\documentclass[11pt]{amsart}

\usepackage{amsmath}
\usepackage{amssymb}
\usepackage{amsthm}
\usepackage{mathrsfs}

\usepackage{slashed}
\usepackage{color}
\usepackage{esint}

\usepackage{graphicx}
\usepackage[all,cmtip]{xy}
\usepackage{hyperref}

\usepackage{epigraph}

\usepackage{fancyhdr}
 
\usepackage{etoolbox}

\makeatletter
\patchcmd{\epigraph}{\@epitext{#1}}{\itshape\@epitext{#1}}{}{}
\makeatother

\newtheorem{thm}{Theorem}%[section]

\newtheorem{prop}{Proposition}

\newtheorem{defn}{Definition}
\newtheorem{examp}{Example}

\hypersetup{
    pdfborder = {0 0 0},%
    colorlinks,%
    citecolor=blue,%
    filecolor=black,%
    linkcolor=red,%
    urlcolor=green
}

\begin{document}

\title{Index theory and noncommutative geometry of topological insulators}

\author{Dan Li}
\address{Department of Mathematics, Purdue University, 150 N. University St, West Lafayette, IN, USA 47907}
	 \email{li1863@math.purdue.edu}

%\date{\today}
 
%\keywords{}

\begin{abstract}
 In this chapter, we report the recent progress in the understanding of the rich mathematical structures of topological insulators in the framework of index theory and noncommutative geometry. 
   
   Topological insulators are new materials observed in nature which behave like insulators in the bulk but have conducting edge  states on the boundary.
   In physics, topological insulators are gapped electronic systems which show topologically protected non-trivial phases in the presence of the time reversal $\mathbb{Z}_2$-symmetry.
   Because of the time reversal symmetry, topological insulators are characterized by a   $\mathbb{Z}_2$-valued invariant, the so-called  topological $\mathbb{Z}_2$ invariant.
   
   One main task is to understand the topological $\mathbb{Z}_2$ invariant in a conceptual and computational framework. 
   The topological $\mathbb{Z}_2$ invariant counts the parity of Majorana zero modes (or Dirac cones), which are quasi-particles similar to Majorana spinors. In quantum field theory, the topological $\mathbb{Z}_2$ invariant is a global parity anomaly, 
   which can be translated into a local gauge anomaly. The relevant gauge problem can be solved completely by index theory. 
   In other words, the topological $\mathbb{Z}_2$ invariant can be understood as a mod 2 version of the Atiyah--Singer index theorem. 
   In a modern language, the topological $\mathbb{Z}_2$ invariant can be obtained by the index pairing between K-homology and K-theory.

   Besides the conceptual understanding of the topological $\mathbb{Z}_2$ invariant, another task is to study the bulk-boundary correspondence, which is an instance of the holographic principle in condensed matter physics. 
   The bulk is given by the momentum space, and our novel observation is that the effective boundary can be identified as the fixed points of the time reversal symmetry. If the bulk and boundary theories are modeled by K-theories,
   then the bulk-boundary correspondence is supposed to be a map connecting these two K-theories, i.e., a KK-cycle in the bivariant K-theory.

   Noncommutative geometry has been developed to be a unified framework for investigating quantum systems and index problems  based on operator algebras, which provides a useful approach to study  disordered topological insulators.
   We  discuss about a noncommutative generalization of the index theory of topological insulators, and propose a new bulk-boundary correspondence using fixed point algebra. 
\end{abstract}

\maketitle

\tableofcontents

\epigraph{``Symmetry is a vast subject, significant in art and nature. Mathematics lies at its root, and it would be hard to find a better one on which to demonstrate the working of the mathematical intellect.'' }
  {--- \textup{Hermann Weyl}, Symmetry}

\section{Introduction} \label{intro}
 Symmetry plays an important role in modern physics and mathematics \cite{W52}.
 In the framework of symmetry protected topological (SPT) orders \cite{CGLW13}, topological insulators are  referred to as $U(1)$ (charge conservation) and
 time reversal symmetry protected topological order. The goal of this chapter is to understand the geometry and topology of  time reversal symmetry in topological insulators.

 Due to time reversal symmetry, topological insulators can be characterized by a $\mathbb{Z}_2$-valued  invariant, which is called the topological $\mathbb{Z}_2$ invariant. 
 If this $\mathbb{Z}_2$  invariant originates from  K-theory, then one expects an index theorem behind it, since  K-theory is the natural receptacle of an index number. 
 As a milestone in mathematics and physics, the Atiyah--Singer index theorem  states
 that  the analytical index is equal to the topological index for an elliptic  operator on a compact manifold.  In physics,
 index theory can be used to understand quantum anomalies such as the chiral anomaly. As for topological insulators, 
 the topological $\mathbb{Z}_2$ invariant is a parity anomaly, which will be understood as a gauge anomaly via index theory. 
 
 This chapter is intended for graduate students interested in the index theory of time reversal invariant topological insulators. 
 The materials in this chapter are not meant to be complete or fully rigorous, rather their goal is to motivate casual readers to
 learn more about the subjects discussed here by consulting the literature. 
 The geometry  of topological insulators is an active research topic in mathematical physics.

\section{Topological band theory} \label{secTopband}

 Time reversal symmetry is the $\mathbb{Z}_2$-symmetry that reverses the direction of physical time. 
 In classical physics, a variable is either even or odd with respect to  the time reversal symmetry.
 For example, position and energy are even, and momentum is odd. 
 In quantum mechanics, time reversal symmetry is represented by an anti-unitary operator $\Theta$, called the time reversal operator. 
 For a system with half-integer total spin, time reversal symmetry has the important property $\Theta^2 = -1$, which gives rise to the Kramers degeneracy, 
 that is, there exists a twofold degeneracy in each band. More precisely, if $\phi$ is an electronic state, then $\Theta \phi$ is a partner state with the same energy as $\phi$ 
 and the pair $(\phi, \Theta \phi)$ is called a Kramers pair. 
 
 A topological insulator is a gapped  fermionic  system, 
 the Fermi energy level is assumed to be located in the band gap.  Furthermore, the Fermi energy is defined
 as the zero energy, gapless edge states (between the conduction and valence bands) may pass through the zero energy level   and intersect with each other. 
 The fundamental objects to study in topological insulators are Majorana zero modes. Physically,
 Majorana zero modes are quasi-particle excitations bound to a defect at
 zero energy. For example, Dirac cones are Majorana zero modes in three-dimensional  time reversal invariant fermionic systems. 
 Two Majorana zero modes tend to couple together and behave as an effective composite boson,   the presence of an unpaired Majorana zero mode
 is the characteristic of a non-trivial topological insulator. 
 We have to point out that Majorana zero modes should not be confused with Majorana fermions (or spinors).  

 One is interested in the band structure of a topological insulator, and the given raw data is a single-particle Hamiltonian.
 On the one hand, one considers the number of Majorana zero modes, which gives rise to the analytical index. 
 Indeed, the parity of Majorana zero modes defines the topological $\mathbb{Z}_2$  invariant. On the other
 hand, one models the band structure by a topological vector bundle 
 and obtains an element  in topological K-theory. The transition function of such a vector
 bundle describes a gauge transformation induced by the time reversal symmetry, and it can be used to compute the topological $\mathbb{Z}_2$ invariant.
 In this section,  we will  look at the topological band theory in the presence of time reversal symmetry, 
 and  give a definition of the topological $\mathbb{Z}_2$ invariant.    
 
 In condensed matter physics, one starts with a lattice model  and investigates the effective Hamiltonian defined over the lattice
 in order to understand different phases of matter and possible phase transitions. 
 It is convenient to consider a continuous space rather than the discrete lattice, and
 the momentum space (or $\mathbf{k}$-space) is used as the base space. Let $X$ be a compact space
  representing the momentum space, 
 for instance, the torus $\mathbb{T}^d$ and the sphere $\mathbb{S}^d$ give two basic examples.

 An involutive space $(X, \tau)$ is a compact space $X$ equipped with an involution $\tau: X \rightarrow X$
 such that $\tau^2 = id_X$. The pair $(X, \tau)$ is also called a Real space with the real structure $\tau$.  
 A typical example is $ \mathbb{R}^{p,q} : = \mathbb{R}^{p} \oplus i \mathbb{R}^{q}$,  the real structure $\tau$ on $\mathbb{R}^{p,q}$ is defined by the complex conjugation, 
 i.e., $\tau : (x, y) \rightarrow (x, -y)$, or equivalently $\tau|_{\mathbb{R}^{p}} = 1$ and  $\tau|_{i \mathbb{R}^{q}} = -1$.

Time reversal symmetry defines an involutive time reversal transformation that flips the sign of a local coordinate.
The explicit form of time reversal transformation may vary according to the choice of coordinate system.
\begin{defn}
 Time reversal symmetry defines the time reversal transformation on the momentum space $X$,
 \begin{equation}
   \tau: X \rightarrow X; \quad   {x} \mapsto -{x}
 \end{equation}
so that $(X, \tau)$ is a Real space.
\end{defn}

\begin{examp}
   The unit sphere $\mathbb{S}^d$ in Cartesian coordinates is
   \begin{equation}
        \mathbb{S}^d = \{ (x_0, x_1, \cdots x_d) \in \mathbb{R}^{d+1} ~|~ x_0^2 + x_1^2 + \cdots x_d^2 = 1\}
   \end{equation}
   the time reversal transformation on $\mathbb{S}^d$ is defined by
   \begin{equation}
        \tau: \mathbb{S}^d \rightarrow \mathbb{S}^d; \quad  (x_0, x_1, \cdots x_d) \mapsto  (x_0, -x_1, \cdots, -x_d)
   \end{equation}
   so the unit sphere with this involution is  denoted by $ \mathbb{S}^{1, d} \subset \mathbb{R}^{1, d}$. 
   In the same coordinate system, a torus is written  as $ \mathbb{T}^d = (\mathbb{S}^{1,1})^d$ since  a circle is $\mathbb{T} = \mathbb{S}^{1,1}$.
   
   In contrast, if the torus $\mathbb{T}^d$ is parametrized by angular parameters,
   \begin{equation}
       \mathbb{T}^d = \{ (e^{i\theta_1}, \cdots , e^{i\theta_d}) ~|~ \theta_i \in [-\pi, \pi ]\, mod\,\, 2\pi, i = 1, \cdots d \}
   \end{equation}
   the time reversal transformation in this case is defined by
   \begin{equation}
       \tau:  \mathbb{T}^d \rightarrow \mathbb{T}^d; \quad (e^{i\theta_1}, \cdots , e^{i\theta_d}) \mapsto (e^{-i\theta_1}, \cdots , e^{-i\theta_d})
   \end{equation}

\end{examp}

The fixed points of an involution $\tau$ is the set of points
\begin{equation}
   X^\tau := \{ {x} \in X ~|~ \tau( {x} ) =  {x} \}
\end{equation}
and we  assume that $X^\tau$ is a finite set. $X^\tau$ is also viewed as
the set of real points with respect to the real structure $\tau$.

\begin{examp}
 The unit sphere $\mathbb{S}^d$ has two fixed points under the time reversal transformation, $(\mathbb{S}^d)^\tau = \{ (\pm1, 0, \cdots, 0) \}$.
    The torus $\mathbb{T}^d$ has $2^d$ fixed points under the time reversal transformation,  $(\mathbb{T}^d)^\tau = \{ (\pm1, \pm1, \cdots, \pm1) \}$.
\end{examp}
 
 Suppose the momentum space $X$ is a CW complex so that $X$ can be constructed by gluing together cells of different dimensions.  
 The extra structure  $\tau$ turns  $X$ into a $\mathbb{Z}_2$-CW complex,
 that is, there exists a $\mathbb{Z}_2$-equivariant cellular decomposition of $X$. 
 Starting with the fixed points $X^\tau$, $X$ can be built up by gluing cells that carry a free $\mathbb{Z}_2$ action, i.e., $\mathbb{Z}_2$-cells. 
 \begin{examp}
   In condensed matter physics, the torus $\mathbb{T}^d$ is often called the Brillouin torus and $\mathbb{T}^{d-1} \times I$ is called the effective Brillouin zone,  where $I = [0,1]$ is the closed unit interval. 
   By the $\mathbb{Z}_2$-CW complex structure,  $\mathbb{T}^d$ can be recovered by $\mathbb{T}^{d-1} \times I$ combined with the free $\mathbb{Z}_2$ action.
 \end{examp}

 Let $H(x)$ be a  single-particle Hamiltonian parametrized by $X$, $H$ is time reversal invariant if $\Theta H(x) \Theta^* = H(\tau(x))$ for all $x \in X$. If $\phi$ is an eigenstate satisfying the eigenvalue equation
 $   H(x) \phi(x) = E(x)\phi(x) $, then the partner state $\Theta \phi$ is the reflection of $\phi$  up to a phase $ \Theta \phi(x) = e^{i \beta(x)}\phi(\tau(x))$ 
 and  satisfies a similar eigenvalue equation 
 \begin{equation}
     H(\tau(x))  \Theta \phi(x) = E(\tau(x))\Theta \phi(x) 
 \end{equation}
 In other words, the Kramers pair $(\phi, \Theta \phi)$ has the same band function $E$.
 The band structure of a topological insulator is described by the finite  set of  band functions $\{ E_n: X \rightarrow \mathbb{R} \}$, 
 and there exists a Kramers pair living in each band. 
 The characteristic property of a topological insulator is completely determined by the edge states passing through the Fermi energy,
 so it is enough to consider  the top band.   
 Applying a partial Fourier transform to the bulk Hamiltonian (or eigenstate), one obtains a boundary Hamiltonian (or eigenstate), 
 from now on we forget about such a partial Fourier transform and only consider the bulk Hamiltonian $H$. After this identification, 
 the edge states between the conduction and valence bands are adjusted to the corresponding bulk states in the top band, and the Fermi energy level is also shifted to the middle of the top band. 
 
 In general, a Hilbert bundle is a complex vector bundle equipped with a complete Hermitian metric, which is commonly used to model 
 continuous fields of Hilbert spaces in geometric quantization.
 The band structure of a topological insulator defines a Hilbert bundle $\pi: \mathcal{H} \rightarrow X$, which is  of  rank 2 since we concentrate on the top band by assumption.  
 If an electronic state is described by a line bundle $\pi: \mathcal{L} \rightarrow X $, then the Hilbert bundle  is modeled by
 $  \mathcal{H} = \mathcal{L} \oplus \tau^*\overline{\mathcal{L}}$ inspired by the Kramers pair,
 where $\tau^*\overline{\mathcal{L}} $ is the pullback  of the conjugate line bundle $\overline{\mathcal{L}}$.

 A Real vector bundle $(E, \iota)$ over a Real space $(X, \tau)$ is a complex vector bundle $E$ over $X$  with an
 involution $\iota$ such that $\iota^2 = id_E$, where $\iota$ is an anti-linear bundle isomorphism  $\iota: E \rightarrow \overline{E}$.
 This entails that $\iota$ induces an anti-linear isomorphism of vector spaces between the fibers  $E_x$ over $x$ and   $\overline{E}_{\tau(x)}$ over $\tau(x)$.
 A Quaternionic vector bundle $(E, \chi)$ over   $(X, \tau)$ is a complex vector bundle $E$ over $X$  with an
 anti-linear anti-involution $\chi$  such that $\chi^2 = -id_E$.

\begin{examp}
 Taking  time reversal symmetry into account, the Hilbert bundle  $\pi: \mathcal{H} \rightarrow X$ becomes
   a Quaternionic vector  bundle $\pi: (\mathcal{H}, \Theta) \rightarrow (X, \tau)$, where $\tau$ is the time reversal transformation
   such that $\tau^2 = 1$ and $\Theta$ is  the time reversal operator satisfying $\Theta^2 = - 1$.
\end{examp} 

 The transition function of the Hilbert bundle $\pi: (\mathcal{H}, \Theta) \rightarrow (X, \tau)$ is given by 
 \begin{equation}\label{transfun}
  w : (X, \tau) \rightarrow (U(2), \sigma)
 \end{equation}
 where $\sigma$ is an involution on the structure group $U(2)$ induced by the time reversal symmetry,
 \begin{equation}
    \sigma: U(2) \rightarrow U(2); \quad g \mapsto -g^T, \quad s.t. \,\, \sigma^2 = 1
 \end{equation}
 The compatibility condition between the involutions $\tau$ and $\sigma$ gives an important property of $w$,
 \begin{equation}
     w \circ \tau = \sigma \circ w  \quad \Leftrightarrow \quad  w(\tau(x)) = -w^T(x), \quad \forall \,\, x \in X
 \end{equation}
 where $T$ denotes the transpose of a matrix. In particular, $w$ becomes a skew-symmetric matrix over the fixed points, 
 \begin{equation}
    w (x) = -w^T(x), \quad \forall \,\, x \in X^\tau
 \end{equation}

\begin{examp}
  When $X = \mathbb{S}^3 = \{ (\alpha, \beta) \in \mathbb{C}^2, \, s.t. \, |\alpha|^2 + |\beta|^2 = 1 \}$,
   the time reversal transformation is defined by $\tau (\alpha, \beta) = (\bar{\alpha}, - \beta)$,
  and the fixed points are    $(\alpha, \beta) = (\pm 1, 0)$.
  In addition, the transition function $w$ is given by
  \begin{equation}
    w: \mathbb{S}^3 \rightarrow SU(2); \quad ( \alpha, \beta) \mapsto \begin{pmatrix}
                                                                     \beta & \alpha \\
                                                                     -\bar{\alpha} & \bar{\beta}
                                                                    \end{pmatrix}
  \end{equation}
\end{examp}

 In the literature, the transition function $w$ is also called the sewing matrix \cite{G17}.
 In physical terms, $w$ is the gauge transformation  induced by the time reversal symmetry. 
 This transition function is the key to constructing a local  formula to compute the topological $\mathbb{Z}_2$ invariant. 
 
 Now we are ready to define the topological $\mathbb{Z}_2$ invariant. 
 Physicists originally proposed two $\mathbb{Z}_2$ invariants from different perspectives. The
 Kane--Mele invariant was first introduced in the study of the quantum spin Hall
 effect in graphene \cite{KM0501}, and generalized to 3d topological insulators later \cite{FKM07}.  
 On the other hand,  the Chern--Simons invariant \cite{QHZ08} was defined as (dimensional reductions of) a topological index, 
 which is also called the Wess--Zumino--Witten (WZW) topological term in string theory.  
 These two $\mathbb{Z}_2$ invariants are equivalent, and both will be called the topological $\mathbb{Z}_2$ invariant.
 The definition of the Kane--Mele invariant is given below, and the topological index  will be discussed in detail  later.
  
 \begin{defn}[\cite{KM05}]\label{KMinv}
  The Kane--Mele invariant, denoted by $\nu$, is defined as  
\begin{equation}
   \nu := \prod_{x \in X^\tau} sgn(pf[w(x)])  = \prod_{x \in X^\tau} \frac{pf [w(x)]}{\sqrt{\det [w(x)]}}
\end{equation}
where the restricted transition function $w|_{X^\tau}  : X^\tau \rightarrow U(2)$ is skew-symmetric.
 \end{defn}
Recall the relation between the Pfaffian and determinant functions is given by
$pf^2(A) = \det(A)$  for a skew-symmetric matrix $A$.
By comparing the Pfaffian with the square root of the determinant of $w$ at $x \in X^\tau$, the Kane--Mele invariant is the product of
the signs of Pfaffians over the fixed points. The Kane--Male invariant reflects the effective quantum field theory of a topological insulator, 
whose geometric picture is a comparison of orientations between the determinant line bundle and a Pfaffian line bundle.

When the Kane--Mele invariant $\nu = -1$, the material is a non-trivial topological insulator, and when
$\nu = 1$, the material is a classical insulator.  
\begin{equation}
  \nu  = \begin{cases} 
          +1, \quad \text{classical insulator} \\
          -1, \quad \text{topological insulator}
        \end{cases}
\end{equation}

\section{Topological K-theory} \label{secTopK}
 
 As a generalization of the direct sum of two vector spaces, the Whitney sum (as a fiberwise direct sum)  defines an additive operation between  vector bundles over a compact space $X$.
 With the Whitney sum, the set of isomorphism classes of vector bundles becomes a semigroup (or a commutative monoid). Similar to the construction from the monoid of natural numbers $\mathbb{N}$ to the abelian group 
 of integers $\mathbb{Z}$,  one defines the abelian group $K(X)$, called the K-theory of $X$, by adding formal differences of two bundles. 
 Such a construction from  semigroup to  group  is called the Grothendieck construction, and the resulting abelian group is called a Grothendieck group. 
 In general, $K(X)= KU(X)$ denotes the complex K-theory in the category of complex vector bundles, and $KO(X)$ denotes the real K-theory constructed from  real vector bundles. 
 By definition, topological K-theory classifies stable isomorphism classes of vector bundles over a compact space.
 
 A discrete $\mathbb{Z}_2$ symmetry   introduces a real structure, the relevant K-theory is the Real K-theory first introduced by Atiyah \cite{A66}. 
 The  Real K-theory $KR(X, \tau)$ is defined as the Grothendieck group of Real vector bundles $(E, \iota)$ over a compact Real space $(X, \tau)$. Similarly 
 the Quaternionic K-theory $KQ(X, \tau)$ is defined to be the Grothendieck group of finite rank Quaternionic vector bundles $(E, \chi)$ over  $(X, \tau)$.
 From the previous section, the topological band theory of a topological insulator is described by a Quaternionic vector bundle over the momentum space $\pi: (\mathcal{H}, \Theta) \rightarrow (X, \tau)$.  
 Hence all possible   band structures of a topological insulator can be classified by the KQ-theory $KQ(X, \tau)$.

If the involution $\tau$ is understood in the context, then it is always omitted from the notation, e.g.,  $KR(X) = KR(X, \tau)$.
The trivial Quaternionic vector bundle over a point is given by $\mathbb{C}^2 \rightarrow pt$ with the trivial Quaternionic structure $\chi = \begin{pmatrix} 
                                                                                                                              0 & -1 \\
                                                                                                                              1 & 0
                                                                                                                            \end{pmatrix}$.
The reduced  Real K-group  $\widetilde{KR}(X)$ is defined as the kernel of the restriction map $ i^*: KR(X) \rightarrow KR(pt) = \mathbb{Z}$, where $i$ is the inclusion of a point $i : pt \hookrightarrow X$.
Higher KR-groups are defined by
\begin{equation}
 KR^{-p, -q}(X) := KR(X \times \mathbb{R}^{p, q})
\end{equation}
There exists an isomorphism, the so-called $(1,1)$-periodicity of KR-theory,
\begin{equation}
 KR^{p+1, q+1}(X) \cong KR^{p,q }(X)
\end{equation}
so by convention a KR-group is denoted by $KR^{p-q}(X) =KR^{p,q }(X) $. The Bott periodicity of KR-theory is 8,
\begin{equation}
 KR^{n+ 8 }(X) \cong KR^{n}(X)
\end{equation}
There exists a canonical isomorphism between KQ-theory and KR-theory,
\begin{equation}
 KQ^n(X) \cong KR^{n\pm4}(X)
\end{equation}
Even though the topological band theory of a topological insulator is classified by KQ-theory,
in practice, we tend to use KR-theory due to the developed computational tools in mathematics. 
 
When the involution $\tau$ on the base manifold $X$ is trivial, notice that $\tau$ is obviously trivial on $X^\tau$, KR-theory  becomes  KO-theory, 
\begin{equation}
 KR^n(X, \tau = id) = KO^n(X)
\end{equation}
The Bott periodicity of KO-theory is 8, i.e., $KO^{n}(X) \cong KO^{n+8}(X)$.
 For instance, the KO-theory of a point  is given by the following table.
\begin{center}
 \begin{tabular}{||c| c| c| c | c | c | c | c |c |c |c ||}
 \hline
   $i$ & 0 & 1 & 2 &  3 & 4 & 5 & 6 & 7  \\ [0.5ex]
 \hline\hline
 $KO^{-i}(pt)$ & $\mathbb{Z}$ & $\mathbb{Z}_2$ & $\mathbb{Z}_2$   & 0 & $\mathbb{Z}$  & 0  & 0 & 0 \\ [1ex]
 \hline
\end{tabular}
\end{center} 

\begin{examp} The KQ-theory of a sphere or a torus may be computed by iterative decomposition into  $\mathbb{R}^{0,k}$  and points, e.g.,
$$
 KQ(\mathbb{S}^{1,2}) = KR^{-4}(\mathbb{S}^{1,2}) =  KO^{-4}(pt) \oplus  KO^{-2}(pt) =  \mathbb{Z} \oplus \mathbb{Z}_2
$$
$$
KQ(\mathbb{T}^2) = KR^{-4}(\mathbb{T}^2) =  KO^{-4}(pt) \oplus  KO^{-2}(pt) =  \mathbb{Z} \oplus \mathbb{Z}_2
$$
$$
KQ(\mathbb{S}^{1,3}) = KR^{-4}(\mathbb{S}^{1,3}) =  KO^{-4}(pt) \oplus  KO^{-1}(pt) =  \mathbb{Z} \oplus \mathbb{Z}_2
$$
$$
KQ(\mathbb{T}^3) = KR^{-4}(\mathbb{T}^3) =  KO^{-4}(pt) \oplus  3 KO^{-2}(pt)  \oplus KO^{-1}(pt)=  \mathbb{Z} \oplus 4 \mathbb{Z}_2
$$
\end{examp}

In the above example, each KQ-theory has at least one $\mathbb{Z}_2$ component, and we know the topological $\mathbb{Z}_2$ invariant belongs to $KQ(X)$. 
For instance, $KQ(\mathbb{T}^3)$ has $\mathbb{Z}_2$ components from both $KO^{-1}(pt)$ and $KO^{-2}(pt)$. So it is natural to
ask where does the topological $\mathbb{Z}_2$ invariant really live in, $KO^{-1}(pt)$ or $KO^{-2}(pt)$? In order to answer this question, one has to
look into the local picture of Majorana zero modes. Later index theory will tell us that the topological
$\mathbb{Z}_2$ invariant really lives in $KO^{-2}(pt)$.

 If one considers the spectral flow of a family of self-adjoint operators along a path (or a loop),
 then the relevant K-theory is $K(X \times S^1) = K^{-1}(X)$ instead of $K(X)$, the same argument holds for KQ-theory. 
 This is the very trick behind the transition from $KQ(X)$ to $KQ^{-1}(X)$. 
 It is very important to distinguish the even KQ-group $KQ(X)$ from the odd $KQ^{-1}(X)$, since the topological index formulas are totally different for the even and odd cases.  
 
 \begin{examp} For an odd-dimensional space $X$, it is natural to consider the odd KQ-group $KQ^{-1}(X)$, 
 which is intimately related to the odd topological index of a gauge transformation \cite{G93}. 
 For instance,  for a 3d sphere or torus, $KQ^{-1}(X)$ leads to the desired $KO^{-2}(pt)$, which is the receptacle of the topological $\mathbb{Z}_2$ invariant.
 $$
 KQ^{-1}(\mathbb{S}^{1,3})= KR^{-5}(\mathbb{S}^{1,3}) =  KO^{-5} (pt) \oplus KO^{-2}(pt)=  \mathbb{Z}_2 
$$
$$
 KQ^{-1}(\mathbb{T}^3)= KR^{-5}(\mathbb{T}^3) = 3 KO^{-4} (pt) \oplus KO^{-2}(pt)= 3  \mathbb{Z} \oplus \mathbb{Z}_2 
$$
 \end{examp}

 K-theory is the natural receptacle of index theory, since the index bundle of a family of Fredholm operators gives a prototypical example of a virtual vector bundle. 
 On the other hand, spin geometry provides a beautiful framework for studying the geometry of spinors such as Dirac, Majorana and Weyl fermions.
 The geometry of Dirac spinors, i.e., Dirac geometry, produces elegant theorems such as the index theory of Dirac operators. 
 It is via Dirac geometry that topological K-theory connects to spin geometry  and index theory.

 In general, a Fredholm operator $F$ is a bounded linear operator with finite-dimensional kernel $ker F$ and cokernel $coker F$, and with closed range $ran F$.
 Given a Fredholm operator $F$,  one defines the Fredholm index as the analytical index,  
\begin{equation}
 ind_a(F) = \dim ker F - \dim coker F 
\end{equation}
 If $F$ is a family of Fredholm operators parametrized by $X$, then one considers the virtual vector bundle $ker F \ominus coker F$ over $X$, which is called the index bundle.
 The index bundle of a Dirac operator (parametrized by $X$) $D = \begin{pmatrix}
                                                                 0 & D_- \\
                                                                 D_+ & 0
                                                                \end{pmatrix}$ 
                                                                gives a class $[kerD_+] - [kerD_-]$ in $K(X)$.

 We are interested in new geometry of Majorana zero modes compared to Dirac geometry. 
 Depending on the given symmetry, an effective Hamiltonian falls into a subclass of Fredholm operators.
 Analogously, the index bundle of a  time reversal invariant Hamiltonian $H(x)$, $x\in X$ is expected to give rise to a class in $KQ(X)$. 
 However, things are more complicated when the time reversal symmetry is present, now the input data is the pair $(H, \Theta)$.
 Similar to the canonical decomposition of the Dirac operator $D$ based on a spinor bundle, we define an effective Hamiltonian as follows. 
 \begin{defn}
   The effective Hamiltonian of a topological insulator is defined by
    \begin{equation}
              \tilde{H} := \begin{pmatrix}
                             0 & \Theta H \Theta^* \\
                             H & 0
                           \end{pmatrix}
    \end{equation}
   %acting on a Kramers pair $(\phi, \Theta \phi)$. 
 \end{defn}
  The formal difference $[ker H] - [ker \Theta H \Theta^*] $ in K-theory is supposed to describe Majorana zero modes, but as a number $\dim ker H - \dim ker \Theta H \Theta^* $
  always vanishes, since a pair of zero modes $(\phi_0, \Theta \phi_0)$ has the same energy.   
  Hence a localization procedure is needed to look closer at the local picture of Majorana zero modes, and obtain the topological $\mathbb{Z}_2$ invariant as an index.

\section{Majorana zero modes and analytical index}\label{secAanInd} 

We introduce the notion of   Majorana zero modes and show that one interpretation  of the $\mathbb{Z}_2$ invariant is given by
the parity of Majorana zero modes. The Majorana zero modes are similar to those found in Bogoliubov--de
Gennes (BdG) superconductors with particle-hole symmetry.  
In physics, Majorana zero modes \cite{DFN15} are quasi-particle excitations bound to a defect at zero energy. 
Two fermionic  Majorana zero modes tend to  couple together and behave effectively as a boson.
For example, Dirac cones are Majorana zero modes in 3d time reversal invariant fermionic systems,
and the presence of an unpaired Dirac cone is the characteristic  of a 3d non-trivial topological insulator.
 Majorana zero modes are different from Majorana fermions, which have been understood
by the spin representations and spin geometry \cite{LM90}.
 Majorana zero modes  have a new  geometry beyond spin geometry compared to Majorana fermions.

  Let us look at the geometry of Majorana zero modes, and explain how to get the mod 2 analytical index  by localization.
  Let $H$ be a time reversal invariant single-particle Hamiltonian parametrized by $X$ 
such that 
\begin{equation}
 \Theta H(x) \Theta^{*} = H(\tau(x)), \quad \forall \,\, x \in X
\end{equation}  
where $\Theta$ is the time reversal operator.
The effective Hamiltonian of a (free  fermionic) topological insulator is defined by
   \begin{equation}
      \tilde{H}(x) := \begin{pmatrix}
                     0 & \Theta H(x) \Theta^* \\
                     H(x) & 0
                  \end{pmatrix}
   \end{equation}  
$\tilde{H}$ is constructed to describe a quasi-particle consisting of a real particle (modeled by $H$) and its mirror image under time reversal symmetry (modeled by $\Theta H \Theta^*$).
 In spin geometry, a Dirac operator is an intertwining operator between spinors, commonly written as $D = \begin{pmatrix}
                                                                 0 & D_- \\
                                                                 D_+ & 0
                                                                \end{pmatrix}$ with $D_- = D_+^*$.
Similar to a Dirac operator, 
   $\tilde{H}$ is constructed  off-diagonally since a topological insulator is a fermionic chiral system and time reversal symmetry changes chirality. 
  Our definition of the effective Hamiltonian is different from the convention used in physics, where a Hamiltonian is always diagonal such as $  \begin{pmatrix}
                                                                 H_\uparrow & 0 \\
                                                                 0 & H_\downarrow
                                                                \end{pmatrix}$ 
                                                                with $ H_\uparrow$, $H_\downarrow$ modeling spinors with opposite chirality.                                                                                            
   The main reason behind our definition is that we want to construct a skew-adjoint operator, whose analytical index is $\mathbb{Z}_2$-valued.
   Notice that the effective Hamiltonian $\tilde{H}$ is also time reversal invariant, % in the following sense,
   \begin{equation}
\Theta \tilde{H}(x) \Theta^{*} = \begin{pmatrix}
                                   0 & \Theta H(\tau(x)) \Theta^* \\
                                   \Theta H(x) \Theta^* & 0
                                 \end{pmatrix} = \begin{pmatrix}
                                                    0 & H(x) \\
                                                    H(\tau(x)) & 0
                                                 \end{pmatrix} = \tilde{H}(\tau(x)) 
\end{equation} 

By assumption, the single-particle Hamiltonian is self-adjoint $H^*(x) = H(x)$, and  we  have 
\begin{equation}
\tilde{H}^*(x) = \begin{pmatrix}
                   0 & H(\tau(x)) \\
                   H(x) & 0
                 \end{pmatrix}^* = \begin{pmatrix}
                                     0 & H(x) \\
                                     H(\tau(x)) & 0
                                   \end{pmatrix} = \tilde{H}(\tau(x))
\end{equation}
which means the effective Hamiltonian $\tilde{H}$ is neither self-adjoint nor skew-adjoint. 
In order to construct a skew-adjoint operator based on $\tilde{H}$, an approximation will be applied to the raw data ${H}$.
As a remark, we have to point out that the real structures defined by the adjoint $*$-operation and the time reversal operator $\Theta$ are different. 
There are so many distinct real structures in topological insulators and superconductors, which is the main source of confusions about the God-given real structure.

 The effective Hamiltonian $\tilde{H}$  acts on a Kramers pair $ (\phi, \Theta \phi) \in \Gamma(X, \mathcal{H})$, which is viewed as a quasi-particle state  $\Phi = (\phi, \Theta \phi)$. 
 A real structure in operator algebra is an anti-linear isometry  $J$ such that $J^2 = \pm 1$.
 The time reversal operator $\Theta$ gives a real structure such that $\Theta^2 = -1$. 
Define a new real structure induced by $\Theta$  as  
\begin{equation}
 \mathcal{J} := \begin{pmatrix}
                  0 & \Theta^* \\
                   \Theta & 0
               \end{pmatrix} = \begin{pmatrix}
                  0 & -\Theta \\
                   \Theta & 0
               \end{pmatrix}
\end{equation}
satisfying $\mathcal{J}^* = \mathcal{J}$ and $\mathcal{J}^2 = 1$.
With respect to $\mathcal{J}$, the Kramers pair $\Phi = (\phi, \Theta \phi)$ is a Majorana state, since $\Phi$ satisfies the real condition $\mathcal{J} \Phi = \Phi $, i.e., 
\begin{equation}
  \begin{pmatrix}
                  0 & \Theta^* \\
                   \Theta & 0
               \end{pmatrix} \begin{pmatrix} 
                              \phi \\
                              \Theta \phi 
                              \end{pmatrix} =  \begin{pmatrix} 
                              \phi \\
                              \Theta \phi 
                              \end{pmatrix}
\end{equation}

   Our definition of Majorana states is a generalized version of the definition used in physics.
   For example from \cite{DFN15}, Majorana states (or Majorana zero modes) are defined as self-adjoint fermionic operators commuting with the Hamiltonian (in the CAR algebra).
   In our definition, we follow the conventions used in the Dirac geometry, see Example \ref{DiracGeo}. So  the dagger $\dagger$-operation  (commonly used in physics) is
   replaced by a   real structure $J$, and in order for a state $\phi$ to be Majorana (or real), the   condition   $\phi^\dagger = \phi$ is generalized to the real condition
   $J\phi = \phi$.

Our notation of Majorana states can be translated to the familiar Majorana operators commonly used in physics, and vice versa.
Suppose $\gamma_1$ and $\gamma_2$ are Majorana operators,  $f_+$ and $f_-$ are fermionic (annihilation and creation) operators related to $\gamma_i$ by 
\begin{equation}
 f_+ = \gamma_1 + i\gamma_2, \quad f_- = \gamma_1 - i\gamma_2
\end{equation}
Or equivalently, the Majorana operators can be expressed in terms of fermionic operators,
\begin{equation}
 \gamma_1 = \frac{1}{2} (f_+ + f_-), \quad  \gamma_2 = \frac{1}{2i} (f_+ - f_-)
\end{equation}
There exists a state-operator correspondence involved in the discussions here. 
By a Bogoliubov transformation, a Majorana state $\Phi = (\phi, \Theta \phi)$ can be  written as
\begin{equation}
 \begin{pmatrix} 
 \gamma_1 \\
 \gamma_2
 \end{pmatrix} = \begin{pmatrix}
                 1 & i \\
                 -i & 1
                \end{pmatrix} \begin{pmatrix}
                               \phi \\
                               \Theta \phi
                               \end{pmatrix} = \begin{pmatrix}
                                                \phi + i \Theta \phi \\
                                                -i \phi + \Theta \phi
                                                \end{pmatrix}
\end{equation}
The matrix $B = \begin{pmatrix}
                 1 & i \\
                 -i & 1 
                \end{pmatrix}$ induces  the Bogoliubov transformation, in this context, the Majorana states $\gamma_i$ are also called Bogoliubov quasi-particles.
 The fermionic states can be recovered as 
 \begin{equation}
   \tilde{f}_+ = \gamma_1 + \Theta \gamma_2 = 2i \Theta \phi, \quad \tilde{f}_- = \gamma_1 - \Theta \gamma_2 = 2 \phi                
 \end{equation}
 Notice that the real structure used here is  the time reversal operator $\Theta$ instead of $i$ in the original case. 
 The Bogoliubov quasi-particles are equivalently obtained by fermionic states,
 \begin{equation}
    \gamma_1 = \frac{1}{2}(\tilde{f}_+ + \tilde{f}_-), \quad  \gamma_2 = -\frac{\Theta}{2}(\tilde{f}_+ - \tilde{f}_-)
 \end{equation}
 Notice that the imaginary unit $i$ in $\gamma_1 =  \phi + i \Theta \phi$ is a formal symbol to connect two particles and form a quasi-particle. 
 In order to avoid unnecessary confusions   introduced by $i$ (as another real structure), we tend to use the vector form  $\Phi = (\phi, \Theta \phi)$ to denote a Majorana state. 
 
 If a Kramers pair $(\phi, \Theta \phi)$ is written as a Bogoliubov quasi-particle $\gamma_1 = \phi + i\Theta \phi$, then the effective Hamiltonian $\tilde{H} = \begin{pmatrix}
                                                                                                                                                       0 & \Theta H \Theta^* \\
                                                                                                                                                       H & 0
                                                                                                                                                      \end{pmatrix}$ 
                           is accordingly written as $H_1 = H + i \Theta H \Theta^*$. 
 Similarly, for the other Bogoliubov quasi-particle $\gamma_2 = \Theta \phi -i \phi  $, the corresponding Hamiltonian is $H_2 = \Theta H_1 \Theta^* = \Theta H \Theta^* - iH$. 
 The multiplication by $\Theta$ maps $(\phi, \Theta \phi)$ to $(\Theta \phi, -\phi)$, so $\gamma_2$ is the mirror image of $\gamma_1$ under the time reversal symmetry, i.e., $\gamma_2 = \Theta \gamma_1$. 
 Define a new Hamiltonian acting on Bogoliubov quasi-particles $(\gamma_1, \gamma_2) = (\gamma_1, \Theta \gamma_1)$,
 \begin{equation}
 \hat{H} := \begin{pmatrix} 
            0 & H_2 \\
            H_1 & 0
           \end{pmatrix} = \begin{pmatrix}
                           0 & \Theta H_1 \Theta^* \\
                           H_1 & 0
                         \end{pmatrix}
                           = \begin{pmatrix}
                            0 &  \Theta H \Theta^* - iH \\
                           H + i \Theta H \Theta^* & 0 
                           \end{pmatrix}  
 \end{equation}
  At the first glance, the analytical index of $\hat{H}$ is 
  \begin{equation}
    ind(\hat{H}) = dim_\mathbb{H} ker ( H + i \Theta H \Theta^*) - dim_\mathbb{H} ker (\Theta H \Theta^* - iH) = 0
  \end{equation}
  which is always zero since $\Theta H \Theta^* - iH = -i(H + i \Theta H \Theta^* ) $ and $H_1, H_2$ have the same kernel. 
  Therefore, the non-trivial invariant  is expected to be defined as a mod 2 version of the chiral part,
  \begin{equation}
     ind(\hat{H}) := dim_\mathbb{H} ker ( H + i \Theta H \Theta^*)  \,\, \text{ mod 2} 
  \end{equation}
  which is the parity of zero modes of Bogoliubov quasi-particles.
  This heuristic reasoning leads to the desired definition of the topological $\mathbb{Z}_2$ invariant, but we still need a skew-adjoint operator to define a mod 2 analytical index.

 A Bogoliubov quasi-particle $\phi + i \Theta \phi$ is the same as a Majorana state (or Kramers pair) $\Phi = (\phi, \Theta\phi)$. 
 Zero modes of a Majorana state can only be found around the set of fixed points, since chiral states $\phi$ and $\Theta \phi$ could pass through the zero energy level only at a fixed point.

Let us look at a pair of localized Majorana zero modes $(\gamma_1^0, \gamma_2^0 = \Theta \gamma^0_1)$ around a fixed point,
\begin{equation*}
 \gamma_1^0 = \begin{pmatrix}
                  \phi_0 \\
                  \Theta \phi_0
                 \end{pmatrix}, \quad
     \gamma_2^0 = \begin{pmatrix}
                \Theta \phi_0 \\
                - \phi_0 
                \end{pmatrix}
\end{equation*}
At this fixed point, a winding matrix (or a transition function of the Hilbert bundle) takes one of the following form
\begin{equation*}
  w_1 = \begin{pmatrix}
         0 & -1 \\
         1 & 0
        \end{pmatrix} \quad or \quad  w_2 = \begin{pmatrix}
         0 & 1 \\
         -1 & 0
        \end{pmatrix}
\end{equation*}
A direct matrix-vector multiplication shows that 
\begin{equation*}
  w_1 \gamma_1^0 = -\gamma_2^0, \,\, w_2 \gamma_1^0 = \gamma_2^0; \quad w_1\gamma_2^0 = \gamma_1^0, \,\, w_2\gamma_2^0 = - \gamma_1^0
\end{equation*}
In practice, we mainly focus on the real part  $\Phi_0^+ = (\phi_0, \Theta \phi_0) = \gamma_1^0$, and the imaginary part  (with respect to $\Theta$)
 $\Phi_0^- = ( \Theta \phi_0, - \phi_0) = \gamma_2^0$ is skipped implicitly but can be easily restored.
 Inspired by spin geometry, we are actually interested in a Majorana state that changes sign, i.e. $\gamma_1^0  \rightarrow - \gamma_2^0$, 
 so we define it as a  Majorana zero mode. We stress that a Majorana state that remains the same sign, i.e., $\gamma_1^0  \rightarrow \gamma_2^0$, will not be counted as a Majorana zero mode.

\begin{defn}
   A  Majorana zero mode is defined as a localized Majorana state $\Phi_0 = (\phi_0, \Theta \phi_0)$ in a small neighborhood of a fixed point $x \in X^\tau$ so that $\phi_0(x) = \Theta \phi_0(x) = 0 $
  and changes sign  at that fixed point.
\end{defn}

   The local geometry of a Majorana zero mode (after linearization) is a conical singularity  $V(x^2 + y^2 - z^2)$ in 3d, 
     that is, $\phi_0$ and  $\Theta \phi_0$ intersect with each other at that fixed point and form a Dirac cone. 
     The intersection point (or center) of a Dirac cone is called a Dirac point by physicists.

 \begin{defn}
  The topological $\mathbb{Z}_2$ invariant of a 
topological insulator is defined as the parity of Majorana zero modes.
 \end{defn}

Atiyah and Singer introduced a mod $2$ analytical index of a real skew-adjoint elliptic operator $P$ in   \cite{AS71},
\begin{equation}
  ind_a (P) := \dim_{\mathbb{R}} \ker P \quad \text{mod 2}
\end{equation}

  Near a fixed point $x \in X^\tau$, the single-particle Hamiltonian $H(x) \sim D(x)$ can be approximated by a Dirac operator $D$.  Accordingly,
  the effective Hamiltonian $\tilde{H} \sim \tilde{D}$
  is   approximated by a skew-adjoint operator $\tilde{D}$,
  \begin{equation}
  \tilde{D}(x) := \begin{pmatrix}
               0 & \Theta D(x) \Theta^* \\
               D(x) & 0
              \end{pmatrix} = \begin{pmatrix}
                                0 & -D(x) \\
                                D(x) & 0
                                \end{pmatrix}
  \end{equation}
  The big difference is that $D$ is  self-adjoint $D^* = D$, but $\tilde{D}$ is skew-adjoint $\tilde{D}^* = -\tilde{D}$, so it makes sense to take the 
  mod 2 index of $\tilde{D}$.

\begin{thm} 
  By localization, $\tilde{H}$ can be approximated by the skew-adjoint operator $\tilde{D}$,
  the topological $\mathbb{Z}_2$ invariant $\nu$ can be interpreted as the  mod 2 analytical index of  $\tilde{H}$, 
\begin{equation}
 \nu = ind_a(\tilde{H}) := \dim \ker \tilde{H}  \quad \text{(mod 2)} 
\end{equation}
which can be computed by the spectral flow of  $H$ modulo 2.
\end{thm} 

Notice that we actually take the quaternionic dimension in the above analytical index since a zero mode of $H$ is a complex state,
\begin{equation}
 ind_a(\tilde{H}) = \dim_\mathbb{H} \ker \tilde{H}  \quad \text{mod 2}
\end{equation}
Because of the Kramers degeneracy, a Majorana zero mode consists of two complex zero modes of $H$, that is,
$\dim_\mathbb{H} \ker \tilde{H} = \dim_\mathbb{C} \ker H$. Furthermore, zero modes of $H$ can only be found around a fixed point, where $H$ is effectively approximated by a Dirac operator $D$, so
$\dim_\mathbb{C} \ker H = \dim_\mathbb{C} \ker D$. Putting this together, we have
\begin{equation}
  ind_a(\tilde{H}) \equiv dim_\mathbb{C} \ker D \quad \text{mod 2}
\end{equation}

The classifying spaces of KR-groups are constructed by different subspaces of Fredholm operators in \cite{AS69}, which
are also connected to different discrete $\mathbb{Z}_2$ symmetries of general topological insulators in \cite{GS15}.
In the following, we identify the mod 2 analytical index of $\tilde{H}$ as an element in $KO^{-2}(pt)$.
Let $\mathcal{F} (\mathscr{H}, J)$ denote the space of Fredholm operators  on a Real Hilbert space
$(\mathscr{H}, J)$, where $\mathscr{H}$ is a complex Hilbert space and $J$ is a real structure such that $J^2 =  1$. In addition, let
$\hat{\mathcal{F}} (\mathscr{H}, J)$ denote the subspace of skew-adjoint Fredholm operators,
which is a classifying space of $KR^{-1}$,
 i.e., 
 \begin{equation}
   KR^{-1}(X) = [X, \hat{\mathcal{F}} (\mathscr{H}, J)]
 \end{equation}
 for a compact Real space $(X, \tau)$.
The effective Hamiltonian $\tilde{H}$ is approximated by a skew-adjoint Fredholm operator acting on the
Real  Hilbert space $(L^2(X, {\mathcal{H}}), \mathcal{J})$,
so the analytical index belongs to $KO^{-2}(pt)$,
\begin{equation}
   ind_a:  \hat{\mathcal{F}} (L^2(X, {\mathcal{H}}), \mathcal{J}) \rightarrow  KO^{-2}(pt)  
\end{equation}
Note that if  $\mathscr{H}$ is a complex Hilbert space, then the analytical index  lives in $KO^{-1}(pt)$.
However, in our case  $L^2(X, {\mathcal{H}})$ is a quaternionic Hilbert
space since for a typical section $(\phi, \Theta \phi)$ and any point $x \in X$,  $(\phi(x), \Theta \phi(x)) \in \mathbb{H} \cong \mathbb{C} \oplus \Theta \mathbb{C}$.
Hence for the analytical index we first take the quaternionic dimension of the kernel space, and then mod 2 to get $ind_a(\tilde{H}) \in KO^{-2}(pt)$.

The geometry of Majorana zero modes studies a quasi-particle consisting of a Dirac fermion and its time reversal partner, which is beyond the Dirac geometry but still in the framework of spin geometry. 
Two Dirac fermions are coupled together in a way that the relevant operator is essentially given by a skew-adjoint Fredholm operator by localization. The geometry of Majorana zero modes 
is described by a coupled product of two Dirac geometries, which will be clear when we discuss about KR-cycles. The same strategy is widely used in index theory, where a general elliptic operator is always reduced to a Dirac operator.

\section{Topological index formula} \label{secTopInd}

The Atiyah--Singer index theorem teaches us that the analytical index can be computed by a topological index.
In this section, we will discuss about the mod 2 topological index, which is originally called the Chern--Simons invariant by physicists \cite{QHZ08}. 
The key observation is that the parity anomaly of the topological $\mathbb{Z}_2$ invariant can be translated into a gauge anomaly, 
and the local  formula is the odd topological index of a specific gauge transformation induced by time reversal symmetry.
In fact, the topological index gives the bulk theory of a topological insulator, which is an integral  over the momentum space as the bulk. 

As a prelude, let us first briefly review the integer quantum Hall effect, whose topological invariant is the first Chern number. 
In the quantum Hall effect,  the Hall conductance can be quantized into intergers, which is  called  the TKNN invariant by physicists. 
This $\mathbb{Z}$-valued  invariant can  be understood by topological band theory and topological K-theory. 
The band structure (or Bloch states) of the quantum Hall system defines a complex vector bundle over the momentum space $X$, which is called the Bloch bundle. 
As  a two-dimensional system, the base manifold is usually taken as a 2-sphere $X = \mathbb{S}^2$ or a 2-torus $X = \mathbb{T}^2$.
Without loss of generality, we assume that the Bloch bundle is of rank one.  
As a consequence, the topological band theory is classified by  complex K-theory, and the topological invariant belongs to $K(X)$, e.g., $\widetilde{K}(\mathbb{T}^2) = \widetilde{K}(\mathbb{S}^2)  = \mathbb{Z}$. 
In addition, the topological index used to compute this $\mathbb{Z}$ invariant is the integral of the first Chern character, also called the first Chern number $c_1$,
\begin{equation}
 c_1 = ind_t(p) = \frac{1}{2\pi} \int_X ch_1(p) = \frac{1}{2\pi}\int_X tr(pdpdp)
\end{equation}
where $p$ is a projection (s.t. $p^2 = p = p^*$) representing the Bloch bundle.

\begin{examp} The Hopf fibration  $S^1 \hookrightarrow S^3 \xrightarrow{\ h \, } S^2$ defines the Hopf bundle as 
a principal $U(1)$-bundle over a 2-sphere $\pi: P \rightarrow S^2 $. 
In complex coordinates, the unit  3-sphere $S^3= \left\{(z_1,z_2)\in\mathbb{C}^2 : |z_1|^2 + |z_2|^2 = 1\right\} $, 
the Hopf map is defined by
\begin{equation}
   h: S^3 \rightarrow S^2; \quad (z_1, z_2) \mapsto ( 2z_1 \bar{z}_2, |z_1|^2 - |z_2|^2),
\end{equation}
 The complex coordinates can be changed to real coordinates by the identification,
 \begin{equation}
     (z_1, z_2)  = (x+i y, z+iw )  \longleftrightarrow (x, y, z, w)
 \end{equation}
 
 In real coordinates, 
 $ {S}^2 = \{ (x, y, z)\in \mathbb{R}^3 \,\, | \,\, x^2 + y^2 + z^2 = 1 \}$.
The Hopf bundle   can be represented by a projection,  
\begin{equation}
p: {S}^2 \rightarrow M_2(\mathbb{C}); \quad 
p(x, y,z) = \frac{1}{2} \begin{pmatrix}
                   1 + z & x + iy \\
                   x- iy & 1-z
                \end{pmatrix}
\end{equation}
The first Chern character 
\begin{equation}
 ch_1(p) = tr(pdpdp) = \frac{-i}{2} ( xdydz - ydxdz + zdxdy)
\end{equation}
is the standard volume form on $ {S}^2$, 
so the first Chern number  $c_1 = 1$.

The  canonical line bundle $\mathcal{L}_H = P \times_{U(1)} \mathbb{C} \rightarrow S^2$ is the  associated line bundle to the Hopf bundle (as a principal $U(1)$-bundle). 
Because of $c_1 = 1$, the first Chern class $c_1(\mathcal{L}_H ) $ gives the generator $\gamma$ of the cohomology group $H^2(S^2 , \mathbb{Z}) = \mathbb{Z} [\gamma ]$. 
The canonical line bundle $\mathcal{L}_H \rightarrow S^2$ satisfies the relation 
\begin{equation}
 \mathcal{L}_H \oplus \mathcal{L}_H  \cong (\mathcal{L}_H \otimes \mathcal{L}_H) \oplus 1_{S^2}
\end{equation}
where $1_{S^2}$ is the trivial line bundle over $S^2$. In the  K-theory ring $K^*(S^2)$, the above relation can be rewritten as
\begin{equation}
 2 [\mathcal{L}_H ] = [\mathcal{L}_H]^2 + [1] \quad \Longleftrightarrow \quad ([\mathcal{L}_H] - [1])^2 = 0
\end{equation}
The relationship between K-theory and reduced K-theory is  $K(X) = \widetilde{K}(X) \oplus \mathbb{Z}$. If one defines the Bott element as 
\begin{equation}
 \beta : = [\mathcal{L}_H] - [1]
\end{equation}
then $\beta$ is the generator of the reduced K-theory $ \widetilde{K}(S^2) = \mathbb{Z} [\beta]$. 

Consider a  magnetic monopole with a unit magnetic charge placed at the origin in $\mathbb{R}^3$, the space $\mathbb{R}^3 \setminus \{ 0 \}$ has the same homopoty type as $S^2$, 
and the corresponding electromagnetic field is modeled by the canonical line bundle  $\mathcal{L}_H \rightarrow S^2$. The first Chern number $c_1 = 1$ confirms the fact that the magnetic monopole
has a unit charge. Indeed, an electromagnetic field can be expressed by the vector potential $A$ as  1-forms in de-Rham cohomology, 
and the field strength $F = dA$ is basically given by the first Chern character, then the integral of $F$ gives us a charge $Q$.

Laughlin argument told us that the quantization of Hall conductance is a result of the gauge invariance. 
In the present geometry,  a charge pump can be applied  to change the magnetic charge, then indirectly change the Hall conductance via the Faraday's law.
If a magnetic flux is pumped adiabatically from infinity to the origin, which induces a gauge transformation 
\begin{equation}
 g: A \mapsto A' = A + \Delta A; \quad F \mapsto F' = F + \Delta F 
\end{equation} 
then the total charge now is the integral of $F'$, and 
the net change of charges $\Delta Q$ is  the integral of  $\Delta F$  (or roughly the gauge transformation $g$) \cite{ASS94}.

\end{examp}

 From integer quantum Hall effect (with a 2d momentum space), one learned  that the relevant topological invariant is   the first Chern number.
 For  a quantum system with time reversal symmetry, it is natural  to consider  a 4d momentum space  
 and expect that the second Chern number would give us the right topological invariant. 
 So for a real material with time reversal symmetry, the relevant topological invariant could be obtained by dimensional reduction from 4d to lower dimensions.
  It is well-known that the first Chern number equals zero for a time reversal invariant system. 

 In four dimensions, the topological index is the integral of the second Chern character $ch_2(F)$, where $F$ is a curvature 2-form. 
 If $c_1 = 0$, then  the second Chern number $c_2$ is the same as the 4d topological index.
 For instance, if the momentum space $X = S^4$,  one decomposes $S^4$ into two hemispheres,
then the topological index can be computed as the winding number over the equator $S^3$,
\begin{equation}
  ind_t = -\frac{1}{8\pi^2}  \int_{S^4} tr (F^2) = \frac{1}{24\pi^2} \int_{S^3} tr(g^{-1}dg)^3
\end{equation}
where $g: S^3 \rightarrow SU(2)$ is a gauge transformation defined on the overlap of two hemispheres (i.e., $S^3$), for details see  \cite{N03}. The right hand side is the winding number of $g$ around
the 3-sphere, i.e., the degree of $g$.

In condensed matter physics,  the non-abelian  Berry  connection and Berry curvature are defined by
\begin{equation}
  a_i = i \langle \psi | \partial_i | \psi \rangle, \quad   f_{ij} = \partial_ia_j - \partial_j a_i + i [a_i, a_j]
  \end{equation} 
As characteristic classes, the second Chern character $ch_2(f) $ and
  the  Chern--Simons 3-form $cs_3(a, f)$ are defined as
  \begin{equation}
     ch_2(f) = tr( f^2 ), \quad cs_3(a, f) =  tr(af - \frac{1}{3}a^3)
  \end{equation}
In fact, $cs_3(a, f)$ is the boundary term of $ch_2(f) $,  
\begin{equation}
   d \,cs_3(a, f) = ch_2(f)
\end{equation}
The magneto-electric polarization $P_3$ over ${\mathbb{T}^3}$ is defined as the Chern--Simons action of the Berry connection,
\begin{equation}
    P_3 (a) := \frac{1}{16\pi^2} \int_{\mathbb{T}^3}  cs_3(a, f)  
\end{equation}
In general, the Chern--Simons action is gauge invariant up to a winding number. More precisely,
under a gauge transformation,
\begin{equation}
  a \mapsto a^g = g^{-1} a g - g^{-1} d g, \quad  g: {\mathbb{T}^3}  \rightarrow U(n)
\end{equation}
one has
\begin{equation}
  \Delta P_3 = P_3(a^g) - P_3(a) = \frac{1}{24 \pi^2} \int_{\mathbb{T}^3}   tr(g^{-1} d g)^3
\end{equation}

The Chern--Simons invariant of a time reversal invariant system is originally defined by the change of magneto-electric polarization under the gauge transformation
induced by the time reversal symmetry modulo two \cite{QHZ08}.
 If  $w$ is the transition function of the Hilbert bundle, then  the Chern--Simons invariant is defined by
\begin{equation}
     \upsilon := \frac{1}{24 \pi^2} \int_{\mathbb{T}^3}    tr(w^{-1} dw)^3 \quad (\text{mod 2})
\end{equation}
One has to take the mod 2 version since the topological invariant of a time reversal invariant system is $\mathbb{Z}_2$-valued from physical observations. 
Indeed, we can prove that the above winding number is naturally $\mathbb{Z}_2$-valued, so ``mod 2'' is redundant.

From the above discussion, the second Chern number $c_2$ over a 4d momentum space is   a universal topological invariant for a time reversal invariant system. 
The 4d topological index is the integral of the second Chern character, which can be reduced to a 3d winding number of a gauge transformation. In topological insulators, 
there is a unique choice for such a gauge transformation, that is, the specific gauge transformation representing time reversal symmetry. Now let us look at the general theory of the 
winding number as a topological index. 

For an odd-dimensional closed manifold $X$, 
the odd Chern character of a differentiable map $g: X \rightarrow U(n)$ (the unitary group) is defined by
\begin{equation}
   Ch(g) := \sum_{k = 0}^{(\dim X - 1)/2} ch_{2k+1}(g) = \sum_{k=0}^{(\dim X - 1)/2} \frac{k!}{(2k+1)!} tr[(g^{-1}dg)^{2k+1}]
\end{equation}
which is a closed differential form of odd degree.
The topological index in odd dimensions is given by the odd index theorem, when the Dirac or A-roof genus
$\hat {A}(X) = 1 $ (e.g., $ \hat{A}(S^d) =\hat {A}(\mathbb{T}^d) = 1  $ ),
\begin{equation}\label{OddIndex}
   ind_t(g)  =      \frac{1}{4\pi^2 }   \int_X Ch(g) %= \frac{1}{4\pi^2 }   \int_X ch_{d} (g), \quad d = \dim(X)
\end{equation} %where $P_F$ is the Fermi projection.
Hence  the winding number (or degree) of $g$ is the same as an odd topological index. 
The above discussion can be rephrased rigorously as the Chern--Weil theory of the relative Chern--Simons form, whose quantization gives the odd index theorem.  

\begin{prop}
  For a 3d topological insulator, the topological $\mathbb{Z}_2$ invariant can be computed by the odd topological index of 
  the transition function representing time reversal symmetry  \eqref{transfun}, i.e.,
  $w : (X , \tau) \rightarrow (U(2), \sigma)$, 
 \begin{equation}
     ind_t(w) = \frac{1}{24 \pi^2} \int_{X}    tr(w^{-1} dw)^3 
\end{equation} 
  which is naturally $\mathbb{Z}_2$-valued.
\end{prop}

 Based on the odd topological index, one can understand why the topological invariant is $\mathbb{Z}_2$-valued as follows. 
 Time reversal symmetry introduces the Kramers degeneracy for an electronic system, e.g., the quantum spin Hall effect. 
 More precisely, each band is doubly degenerate, or there exists a Kramers pair in each band due to the time reversal symmetry.  
 However, there is no a priori ordering between the Kramers pair. 
 If one switches the ordering between the Kramers pair, the transition function $w$ changes sign, and further $ind_t(w)$ changes sign.  
 The winding number $ind_t(w)$ as a global invariant is independent of the choice of orderings, so $ind_t(w)$ must be $\mathbb{Z}_2$-valued 
 to eliminate the ambiguity introduced by the undetermined ordering. As a gauge anomaly, Witten already pointed out that the WZW topological term is $\mathbb{Z}_2$-valued in \cite{W82}.  
 
 The odd topological index formula is also valid  in  one dimensions, e.g.,  the Kitaev Majorana chain. However, the topological index formula   in  two dimensions is more mysterious, 
  the first Chern character modulo $2$ cannot give a local formula for the topological $\mathbb{Z}_2$ invariant since the first Chern number is zero, i.e. $c_1 = 0$. Heuristically,  if one considers the net change of charges $\Delta Q$
  passing through the zero energy level, that is always zero. Indeed, a Majorana zero mode has two opposite-moving chiral states,  one chiral state contributes $+1$ to $\Delta Q$, at the same time the other chiral state contributes $-1$ to $\Delta Q$.
 
Inspired by the relation between   weak and strong topological $\mathbb{Z}_2$ invariants,  we propose a topological index formula below by reducing the dimension from 2d to 1d.
Let $X = \mathbb{T}^2$ be the momentum space (or the Brillouin torus), so $EBZ = \mathbb{S}^{1,1} \times I$  and $\partial EBZ = \mathbb{S}^{1,1}_N \sqcup \mathbb{S}^{1,1}_S $,
where $\mathbb{S}^{1,1}_N, \mathbb{S}^{1,1}_S $ are the north and south circles in the cobordism $\mathbb{S}^{1,1} \times I$. 
Based on the  $\mathbb{Z}_2$-cellular decomposition of $(\mathbb{T}^2, \tau)$, $\mathbb{T}^2$ is decomposed into two copies of  $\mathbb{S}^{1,1} \times I$ and the Hilbert bundle $\pi: \mathcal{H} \rightarrow \mathbb{T}^2$
can be reconstructed from the restricted bundles $\mathcal{H}_{\mathbb{S}^{1,1} \times I}$ by the clutching map $w: \mathbb{S}^{1,1}_N \sqcup \mathbb{S}^{1,1}_S \rightarrow U(2)$.
The topological index formula on $\mathbb{T}^2$ is defined as the product of the 1d indices from the boundaries $\mathbb{S}^{1,1}_N, \mathbb{S}^{1,1}_S$,
\begin{equation}\label{2dTopInd}
   \upsilon(w) = \frac{1}{2\pi}  \int_{\partial (\mathbb{S}^{1,1} \times I)}  tr(w^{-1}dw) :=  [ \frac{1}{2\pi}  \int_{ \mathbb{S}^{1,1}_N}  tr(w^{-1}dw) ][ \frac{1}{2\pi}  \int_{ \mathbb{S}^{1,1}_S}  tr(w^{-1}dw)]
\end{equation}
We use multiplication instead of addition on the right hand side, since the product of two $\mathbb{Z}_2$ is still $\mathbb{Z}_2$ but the addition of two $\mathbb{Z}_2$ is not $\mathbb{Z}_2$. 
\begin{prop}\label{2dtopindex}
    For a 2d topological insulator over $\mathbb{T}^2$, the topological $\mathbb{Z}_2$ invariant can be computed by a variation of the odd  topological index formula 
   \begin{equation}
    ind_t(w) = \frac{1}{2\pi}   \int_{\partial (\mathbb{S}^{1,1} \times I )} tr(w^{-1}dw)
  \end{equation}
 where  $w : {\partial (\mathbb{S}^{1,1} \times I )}  ={\mathbb{S}^{1,1}_N \sqcup \mathbb{S}^{1,1}_S} \rightarrow U(2)$ is the transition function restricted to the boundary of the effective Brillouin torus, 
 and it is naturally $\mathbb{Z}_2$-valued.
\end{prop}

In complex K-theory, the odd K-group $K^{-1}(X)$ has a different interpretation compared to $K(X)$. Roughly, $K^{-1}(X)$ can be defined based on $K(S^1 \times X)$, 
which  is closely related to the spectral flow of a family of self-adjoint operators along a circle.
For a class in an odd K-group, one expects an odd topological index. 
In index theory, the odd topological index is a generalization of the winding number of the determinant function $\det: U(n) \rightarrow \mathbb{C}^\times$, which was first understood as the index of a Toeplitz operator in operator algebra.
In the topological band theory of a topological insulator, the information about the band structure  is essentially encoded in the time reversal gauge $w$,
so the 2d topological index is also reduced to a variation of the odd topological index.
In particular, when $X = \mathbb{T}^2$, $[w] \in KQ^{-1} ({\mathbb{S}^{1,1}_N \sqcup \mathbb{S}^{1,1}_S})$, where ${\mathbb{S}^{1,1}_N \sqcup \mathbb{S}^{1,1}_S}$ is the boundary of the effective Brillouin torus $\mathbb{T} \times I$.
In sum, the local formula of the topological $\mathbb{Z}_2$ invariant is   an odd topological index (or its variation) of the  transition function.
 
\section{Mod 2 index theorem} \label{secIndthm}

The topological $\mathbb{Z}_2$ invariant counts the parity of Majorana zero modes, which is interpreted as an analytical index in \S \ref{secAanInd}. 
Such a parity anomaly is translated into a gauge anomaly, which has an interpretation as an odd topological index in \S \ref{secTopInd}.  
By the Atiyah--Singer index theorem, the analytical index is equal to the topological index through spectral flow. In addition, for a topological insulator,
the analytical index and topological index both are naturally $\mathbb{Z}_2$-valued (indeed $KO^{-2}(pt) = \mathbb{Z}_2$). 

\begin{thm} 
 The  $\mathbb{Z}_2$ invariant of topological insulators  can be understood as a mod 2 index theorem,
 \begin{equation}
    ind_a(\tilde{H}) = ind_t(w)
 \end{equation}
where $\tilde{H}$ is the effective Hamiltonian and $w$ is the time reversal gauge. 
\end{thm} 

There is a famous slogan by Quillen: ``Dirac operators are a quantization of connections, and index theory is a quantization of the Chern character.'' 
If one spells out this slogan for the the odd index theorem, then one obtains the proof of the above theorem. Let us briefly review the ideas involved in the proof in the following paragraph.

Given two connections $ {A}_0$ and $ {A}_1$ on a vector bundle, the relative Chern--Simons form is defined by 
\begin{equation}
   cs(  {A}_0,  {A}_1) : = \int_0^1 tr(\dot{ {A}}_t e^{ {A}_t^2}) dt
\end{equation}
where for $t \in [0,1]$,
\begin{equation}
    {A}_t = (1-t)  {A}_0 + t  {A}_1, \quad \dot{  {A}_t} = \frac{d A_t}{dt} =  {A}_1 -  {A}_0
\end{equation}
There exists a transgression formula connecting the relative Chern--Simons form and the Chern character  in Chern--Weil theory,
\begin{equation}
   d\, cs(  {A}_0,  {A}_1) = ch(  {A}_1 ) - ch(  {A}_0 )
\end{equation}
where the Chern character of a connection $ A$ is defined as  $ch( {A}) := tr(e^{ {A}^2})$. A connection of a vector bundle
has a local form $A = d + \omega$, $d$ is the differential (as the trivial connection) and $\omega \in \Omega^1( \mathfrak{u}(n))$ is a 1-form. 
 For a differentiable map $g : X \rightarrow U(n)$,  choose two connections of the trivial bundle $X \times \mathbb{C}^n $ as 
\begin{equation}
  {A}_0 = d, \quad  {A}_1 =  g^{-1}dg 
\end{equation}
the relative Chern--Simons form $cs(d,  g^{-1}dg)$ gives the odd Chern character by Taylor expansion, 
\begin{equation}
    cs(d,  g^{-1}dg) = Ch(g)
\end{equation}
After quantization, the right hand side gives the odd topological index, and the left hand side is a spectral flow $ sf(D, g^{-1}Dg)$
with the connection $d$ replaced by a Dirac operator $D$.
In our case, the analytical index $ind_a(\tilde{H})$ can be computed by the mod 2 spectral flow of the self-adjoint Hamiltonian $H$, 
and the time reversal gauge $w$ plays the role of the map $g$.
Putting it together, the mod 2 index theorem of a topological insulator can be proved by Quillen's slogan. 

Index theory can be best understood in the framework of K-theory. First, from the discussion about the analytical index in \S \ref{secAanInd},
we  know the topological $\mathbb{Z}_2$ invariant (as an index number) belongs to $KO^{-2}(pt)$. Second, from the topological index  in \S \ref{secTopInd},
the essential information is encoded in the time reversal gauge $w$, which determines a class in $KQ^{-1}(X)$. 
If the starting point is an elliptic operator, then its symbol class gives rise to a class in the K-theory of the cotangent bundle over $X$, i.e., $\pi: T^*X \rightarrow X$. 
In general, the topological index map (constructed by Atiyah) has $T^*X$ as the source and an abstract point (or a Euclidean space $\mathbb{R}^N$) as the target space. 

For a topological insulator, the effective Hamiltonian $\tilde{H}$ can be approximated by a skew-adjoint operator, so 
  the relevant topological index map is 
\begin{equation}
   ind_t : KR^{-2}(T^*X) \rightarrow KO^{-2}(pt)  
\end{equation}
For a $d$-dimensional Real space $(X, \tau)$, the Thom isomorphism in $KR$-theory is given by
\begin{equation}
   KR^{-j}(X) \cong KR^{d-j}(T^*X)
\end{equation}
Combining these two maps gives a map from KR-theory (or KQ-theory) to $KO^{-2}(pt)$, and we still call it the topological index map.

\begin{examp}
\label{indexex}

  When $X = \mathbb{T}^3$, the topological index map is a map from $KQ^{-1}(\mathbb{T}^3) $ to $KO^{-2}(pt)$,
  since the Thom isomorphism identifies $ KQ^{-1}(\mathbb{T}^3) $ with $KR^{-2}(T^*\mathbb{T}^3)$,
   \begin{equation}
        ind_t: KQ^{-1}(\mathbb{T}^3) = KR^{-5}(\mathbb{T}^3)  \cong KR^{-2}(T^*\mathbb{T}^3)  \rightarrow KO^{-2}(pt)  
   \end{equation}
  By iterative decomposition, $KQ^{-1}(\mathbb{T}^3)$ has a component $KO^{-2}(pt) $, and the index map looks like a restriction map (or a localization),
   \begin{equation}
     ind_t: KQ^{-1}(\mathbb{T}^3) = 3 KO^{-4} (pt) \oplus KO^{-2}(pt) \rightarrow KO^{-2}(pt) 
   \end{equation}

   When $X = \mathbb{T}^2$, the topological index map is a map from
   $KQ(\mathbb{T}^2)$ to $KO^{-2}(pt)$, 
   \begin{equation}
        ind_t: KQ(\mathbb{T}^2) = KR^{-4}(\mathbb{T}^2)  \cong KR^{-2}(T^*\mathbb{T}^2) \rightarrow KO^{-2}(pt) 
   \end{equation}
   Or equivalently,
   \begin{equation}
     ind_t: KQ(\mathbb{T}^2) = KO^{-4}(pt) \oplus  KO^{-2}(pt) \rightarrow KO^{-2}(pt) 
   \end{equation}
  By the reduction from $\mathbb{T}^2$ to $\partial(\mathbb{T} \times I) = S^1 \sqcup S^1$, we lift the index map to
    \begin{equation}
       \Delta \circ ind_t:  KQ^{-1}(S^1 \sqcup S^1) \rightarrow  KQ(\mathbb{T}^2)  \rightarrow KO^{-2}(pt) 
   \end{equation}
   where $\Delta$ is a boundary map in KQ-theory.
  Since $ \widetilde{KQ}(\mathbb{T}^2) = KO^{-2}(pt)$  and 
  \begin{equation}
     \widetilde{KQ}^{-1}(S^1) = KQ^{-1}(S^1) = KR^{-5}(S^1) = KO^{-5}(pt) \oplus KO^{-4}(pt)
  \end{equation}
  in the reduced KQ-theory,  one has
     \begin{equation}
      \widetilde{\Delta}:  \widetilde{KQ}^{-1}(S^1 \sqcup S^1) = 2 KO^{-4}(pt)   \rightarrow  \widetilde{KQ}(\mathbb{T}^2) = KO^{-2}(pt) 
   \end{equation}
   which basically is the map 
   \begin{equation}
      \mathbb{Z} \times \mathbb{Z} \rightarrow \mathbb{Z}_2; \quad (m,n ) \mapsto (m\,\, \% \,\, 2 ) \cdot  (n \,\, \% \,\, 2)
   \end{equation}
   where $\%$ is the modulo operation in  arithmetic. 
   We stress that the bulk theory in 2d is given by  $KQ(\mathbb{T}^2)$, so the topological index map is from $ KQ(\mathbb{T}^2)$ to $KO^{-2}(pt) $. However, in order to find a local topological index formula, one is forced to
   reduce from $\mathbb{T}^2$ to  $\partial(\mathbb{T} \times I)$ or lift from $KQ(\mathbb{T}^2) $ to  $KQ^{-1}(S^1 \sqcup S^1)$ and make use of the  class $[w] \in  KQ^{-1}(S^1 \sqcup S^1)$.
\end{examp}

\section{KR-homology and Index pairing}

As discussed in  \S \ref{secTopInd}, the topological index is the integral of the odd Chern character $Ch(g)$ over the momentum space $X$, 
which can be viewed as a pairing between a K-theoretic class $[g] \in K^{-1}(X)$ with the fundamental class $[X]$.   
\begin{equation}
   ind_t(g) = \frac{1}{4\pi^2} \int_X Ch(g) = \langle [X], [g] \rangle 
\end{equation}
In a modern language, an index can be obtained by the index pairing between K-homology and K-theory (as a generalized cohomology theory).
\begin{equation}
   K_*(X) \times K^*(X) \rightarrow \mathbb{Z}; \quad \langle -, - \rangle \mapsto index
\end{equation}
The index pairing between K-homology and K-theory can be generalized to a paring between cyclic homology and cohomology using Connes--Chern characters.
We reformulate the mod 2 index theorem of a topological insulator into an index pairing between KR-homology and KR-theory, 
which is a preparation for the index pairing in noncommutative geometry.

In spin geometry, one models Majorana spinors by $KR$-cycles in $KR$-homology.
As an analogy, we will define a $KQ$-cycle (as a generalized $KR$-cycle) to model  Majorana zero modes. 
Let us first briefly review the definition of a KR-cycle in operator algebra, 
which gives a canonical representative for the  fundamental  class in KR-homology.

  A  K-cycle for a   $*$-algebra of  operators $ \mathcal{A} $ is a triple $(\mathcal{A},\mathscr{H}, D)$,
  also called a spectral triple in noncommutative geometry, where $\mathscr{H}$ is a complex Hilbert space, 
  and $\mathcal{A}$ has a faithful  $*$-representation on $\mathscr{H}$ as bounded operators,
  i.e., $\pi : \mathcal{A} \rightarrow B(\mathscr{H})$. $D$ is a self-adjoint (typically unbounded)  operator satisfying the following conditions. 
  \begin{enumerate}
   \item  $D$ has compact resolvent, that is, for $\lambda \notin \mathbb{R}$, the resolvent $(D - \lambda)^{-1}$ is a compact operator.
   \item    For all $a \in \mathcal{A}$, the commutator $[D, \pi(a)] : = D \pi(a) - \pi(a)D $ is a bounded operator.   
   %\item    For all $a \in \mathcal{A}$, the operator $\pi(a) (1+ D^2)^{-1/2}$ is  compact.  
  \end{enumerate}
  A bounded linear operator $T : \mathscr{H}  \rightarrow  \mathscr{H}$ is called compact if $T$ maps any bounded sequence
 $\{\xi_k \} \in \mathscr{H}$ to a sequence $\{T\xi_k \} \in \mathscr{H}$  with a convergent subsequence. 
 If the representation $\pi$ is understood in the context, it is always  skipped from the notation,
and one treats $\mathcal{A}$ as a subalgebra of $B(\mathscr{H})$  (the algebra of bounded  linear operators with the norm topology).
If the operator $D$ is a Dirac-type operator, then $(\mathcal{A},\mathscr{H}, D)$ is called a Dirac-type spectral triple. 
  
  \begin{examp}
     Conisder the circle $S^1$ parametrized by the angular parameter $\theta$, the canonical K-cycle on $S^1$ is given by $(C^\infty(S^1), L^2(S^1), D)$.
     Here the function algebra is the smooth algebra $C^\infty(S^1)$, the Hilbert space is the square integrable functions $L^2(S^1)$, and the Dirac operator is
     $D = \frac{d}{id\theta}$. 
  \end{examp}

A K-cycle $(\mathcal{A},\mathscr{H}  , D)$ is even (or graded) if there exists a   grading operator $\gamma$ with the properties
 $\gamma^* = \gamma$ and $\gamma^2 = 1$ such that $D\gamma = -\gamma D$ and $\gamma \pi(a) = \pi(a)\gamma$ for all $a \in \mathcal{A}$. 
Otherwise, a K-cycle is odd (or ungraded). If there exists a grading $\gamma$, then the Hilbert space $\mathscr{H} $ is also assumed to be 
$\mathbb{Z}_2$-graded, which is thought of as the space of (square integrable) sections of a spinor bundle.
A typical example of grading operators  is given by $\gamma = \begin{pmatrix}
                                                       1 & 0 \\
                                                       0 & -1
                                                     \end{pmatrix}$.
 
A (general) real structure is defined by an anti-linear isometry $J$ on $\mathscr{H}$. For example, the time reversal symmetry introduces a real structure.

\begin{defn}
  An even  $KR_{2k}$-cycle (depending on $2k$ mod $8$) is defined as an even   K-cycle equipped with a   real structure, i.e., a quintuple $( \mathcal{A}, \mathscr{H}, D, J, \gamma)$,
  satisfying the relations,
  \begin{equation}
      JD = DJ, \quad J^2 = \epsilon, \quad J\gamma = (-1)^k \gamma J
  \end{equation}
  where $J^2 = 1$ if $2k \equiv 0, 6$ mod $8$ and $J^2 = -1$ if $2k \equiv 2, 4$ mod $8$.
\end{defn}

Here we  give the representation theory of spinors in spin geometry \cite{LM90}, also called spin representation. 
By the representation theory of the real Clifford algebra $Cl_{2k, 0}$, when $2k = 0$ mod 8, it has a unique real pinor representation
(i.e., Majorana pinor) and there are two inequivalent real spinor representations (i.e., Majorana--Weyl spinors).
When $2k =2$ mod 8, it has a unique quaternionic pinor representation
(i.e., symplectic Majorana pinor) and there are two inequivalent complex spinor representations (i.e.,  Majorana--Weyl spinors).
When $2k =4$ mod 8, it has a unique quaternionic pinor representation
(i.e., symplectic Majorana pinor) and there are two inequivalent quaternionic spinor representations (i.e.,  symplectic Majorana--Weyl spinors).
When $2k = 6$ mod 8, it has a unique real pinor representation
(i.e., Majorana pinor) and there are two inequivalent complex spinor representations (i.e., Majorana--Weyl spinors).

\begin{examp}
   When $2k \equiv 2$ mod $8$, one has   Majorana--Weyl spinors
   modeled by a $KR_{2}$-cycle $( \mathcal{A}, \mathscr{H}, D,  J, \gamma)$ such that
   \begin{equation}
  JD = DJ, \quad J^2 = -1, \quad J\gamma = - \gamma J
  \end{equation}
\end{examp}

Similarly, when  there is no grading operators, one defines an odd $KR$-cycle by a quadruple $( \mathcal{A},\mathscr{H} , D, J)$.
\begin{defn}
   An odd  $KR_{2k-1}$-cycle (for $2k-1 \equiv 1, 3, 5, 7 \mod 8$) is  defined as a quadruple $( \mathcal{A}, \mathscr{H} , D, J)$ satisfying the relations,
   \begin{equation}
        JD = (-1)^{k} DJ, \quad J^2= \epsilon
   \end{equation}
   where $J^2 = 1$ if $2k-1 \equiv 1, 7$ mod $8$ and $J^2 \equiv -1$ if $2k-1 = 3, 5$ mod $8$.
\end{defn}

The representation theory of the real Clifford algebra in odd dimensions is easier, when $2k-1 = 1, 7$ mod $8$, there is a unique real spinor representation;
when $2k-1 = 3, 5$ mod $8$, there is a unique quaternionic spinor representation.

\begin{examp}
   When $2k-1 = 5$ mod $8$, one has a quaternionic spinor modeled by   a $KR_{5}$-cycle $( \mathcal{A}, \mathscr{H}, D, J)$ such that
   \begin{equation}
     JD = -DJ, \quad J^2= -1
   \end{equation}

\end{examp}

From a K-cycle $(\mathcal{A}, \mathscr{H} , D)$, one obtains the corresponding Fredholm module  $(\mathcal{A}, \mathscr{H} , F)$ by setting $F = D (1+D^2)^{-1/2}$.
By definition, the set of equivalence classes of Fredholm modules modulo unitary equivalence and homotopy equivalence defines the K-homology group, for details see \cite{HR00}.
The following example is the classical Dirac geometry modeling Majorana spinors in spin geometry \cite{LM90, VFG01}.

\begin{examp}\label{DiracGeo}
   Let $M$ be a compact spin manifold of dimension $2k$, its Dirac geometry is defined as the spectral triple
     $(C^\infty(M), L^2(M, \slashed{S}), D\!\!\!\!/ \,)$,
  where
   $L^2(M, \slashed{S})$ is the Hilbert space of spinors and $D\!\!\!\!/$ is the Dirac operator. The grading operator $\gamma$, or $c(\gamma)$ for the Clifford multiplication by $\gamma$,
   is defined as usual in an even dimensional Clifford algebra. In addition, the canonical real structure is the charge conjugation operator $C$ in the Clifford algebra. Thus the quintuple 
   \begin{equation}
    (C^\infty(M), L^2(M, \slashed{S}), D\!\!\!\!/ \, , \gamma, C)
   \end{equation} defines a
   $KR_{2k}$-cycle of spinors. In spin geometry, when a spinor $\psi \in L^2(M, \slashed{S})$ satisfies the real condition $C \psi = \psi$ (similar to $\psi^\dagger = \psi$ in physics), 
   it is called a Majorana spinor, the space of Majorana spinors
   is denoted by $ L^2(M, \slashed{S}; C)$.

\end{examp}

As a summary, a real spectral triple or  KR-cycle satisfies the commutation relations given in the following table. 
\begin{center}
 \begin{tabular}{||c| c| c| c | c | c | c | c |c |c |c ||}
 \hline
   $ j\,\, \text{mod 8}$ & 0 & 1 & 2 &  3 & 4 & 5 & 6 & 7  \\ [0.5ex]
 \hline\hline
  $J^2 = \pm 1$ & $+$ & $+$ & $-$   &  $-$  &  $-$  &  $-$  & $+$ & $+$ \\
 \hline
  $JD = \pm DJ $ & $+$ & $-$ & $+$   &  $+$  &  $+$  &  $-$  & $+$ & $+$ \\  
 \hline
 $J\gamma = \pm \gamma J $ & $+$ &   & $-$   &     &  $+$  &     & $-$ &   \\ [1ex]
 \hline
\end{tabular}
\end{center} 
In addition, a $KR_j$-cycle is said to have the  KO-dimension $j$ mod $8$. KO-dimension of a real spectral triple, first introduced by Connes in \cite{C95}, is
the shift in the grading on KR-homology. The idea comes from the KO-orientation represented by a fundamental class in KO-homology, 
which induces the Poincar{\'e} duality for KO-theory.

Let us construct a generalized KR-cycle to model the geometry of Majorana zero modes. 
Due to the Kramers degeneracy, each band has a Kramers pair with opposite spins. 
The Hilbert bundle is an analogy of a spinor bundle, by assumption, 
\begin{equation}
 \mathcal{H} = \mathcal{L} \oplus \Theta \mathcal{L} \cong \mathcal{L} \oplus \tau^* (\overline{\mathcal{L}})
\end{equation}
In this sense, there exists a natural spin structure on the Hilbert bundle due to the time reversal symmetry. 
So the Hilbert space $L^2(X,  {\mathcal{H}})$ is $\mathbb{Z}_2$-graded,
and it can be decomposed into two chiral components $L^2(X, \mathcal{L}) \oplus  L^2(X, \Theta \mathcal{L})$. If we use the notations 
$\mathscr{H} = L^2(X,  {\mathcal{H}})$ and  $\mathscr{L} = L^2(X, \mathcal{L})$, then 
\begin{equation}
  \mathscr{H} = \mathscr{L} \oplus \Theta \mathscr{L} = \mathscr{H}_+ \oplus \mathscr{H}_-
\end{equation}
is similar to the decomposition of spinors in spin geometry since the time reversal operator changes chirality. 
If we add the usual grading operator
$\gamma = \begin{pmatrix}
          1 & 0 \\
          0 & -1
         \end{pmatrix}$
with $\gamma = \gamma^*$ and $\gamma^2 = 1$, then $\gamma$ can be used to separate the chiral states,
\begin{equation}
 \frac{1 + \gamma}{2} \Phi = \begin{pmatrix}
                               1 & 0 \\
                               0 & 0
                             \end{pmatrix} \begin{pmatrix}
                                             \phi \\
                                             \Theta \phi
                                            \end{pmatrix} = \begin{pmatrix}
                                                               \phi \\
                                                                0
                                                             \end{pmatrix}, \quad
 \frac{1 - \gamma}{2} \Phi = \begin{pmatrix}
                               0 & 0 \\
                               0 & 1
                             \end{pmatrix} \begin{pmatrix}
                                             \phi \\
                                             \Theta \phi
                                            \end{pmatrix} = \begin{pmatrix}
                                                               0 \\
                                                                \Theta \phi
                                                             \end{pmatrix}
\end{equation}

\begin{defn}\label{KQcycle}
 The KQ-cycle of a topological insulator is defined as the quintuple
\begin{equation}\label{KQ6cyc}
   (C^\infty(X), L^2(X, \mathcal{H})  , \tilde{H}, \mathcal{J}, \gamma )
\end{equation}
where
$$
\tilde{H} = \begin{pmatrix}
               0 & \Theta H \Theta^* \\
                H& 0
               \end{pmatrix}, \quad 
                                   \mathcal{J} = \begin{pmatrix}
                                                0 & \Theta^* \\
                                              \Theta & 0
                                                  \end{pmatrix} 
$$
%such that
%$$
%\mathcal{J} \tilde{H} = - \tilde{H}\mathcal{J}, \quad \mathcal{J}^2 = 1, \quad \mathcal{J}\gamma  = - \gamma \mathcal{J}
%$$
\end{defn}

The effective Hamiltonian $\tilde{H}$ is {not} self-adjoint, so $ (C^\infty(X), L^2(X, \mathcal{H})  , \tilde{H})$  is not a K-cycle in the classical sense. 
However, by localization, a  Majorana zero mode consists of an electronic chiral state and its time reversal partner, and each chiral state has a well-defined KR-cycle. 
If we switch off the time reversal symmetry, then a K-cycle  $(C^\infty(X),   L^2(X, \mathcal{H}) ,  {H})$ can be constructed based on the single-particle Hamiltonian $H$. 
Furthermore, if $H$ is approximated by a Dirac operator $D$, then we get the Dirac-type spectral triple  $(C^\infty(X),  L^2(X, \mathcal{H})  ,  D)$. Now let us switch on the time reversal symmetry, 
the quadruple
\begin{equation}
    (C^\infty(X),  L^2(X, \mathcal{H}) , D, \Theta )
\end{equation}
defines an odd $KR_5$-cycle satisfying
\begin{equation}
   \Theta D = - D \Theta,  \quad \Theta^2 = -1 
\end{equation}
Recall that there exists an ambiguity in choosing a chiral state in a Majorana state, so at the same time 
the quadruple
\begin{equation}
    (C^\infty(X),  L^2(X, \mathcal{H}), -D, -\Theta )
\end{equation}
defines  the complementary  $KR_5$-cycle, since $\Theta^* = - \Theta$ and $\Theta D \Theta^* = -D$.
In sum, the localization of the KQ-cycle can be reduced to two $KR_5$-cycles. % so that a Majorana zero mode can be modeled by  a coupled product of two  $KR_5$-cycles.  

 As a result, the geometry of a Majorana zero mode can be modeled by  a coupled product of two  $KR_5$-cycles, which is viewed as a $KR_{10}$-cycle with KO-dimension  $2$.
   Similar to the spin representation when $2k \equiv 2$ mod 8, the $KQ$-cycle models two inequivalent complex spinors $(\phi, \Theta \phi) \in L^2(X, \mathcal{C}) \oplus  L^2(X, \Theta \mathcal{C})$,
   and it can be viewed as a quaternionic pinor $\Phi \in L^2(X,  {\mathcal{H}})$. Hence the localization of the KQ-cycle determines a class in the $KR_2$-homology group, 
   in order to get an index number, it  is expected to be paired with a K-theoretic class in the $KR^{-2}$-group.

   From the discussions in \S \ref{secIndthm},    the topological index map has the source as the $KR^{-2}$-group of the cotangent bundle $\pi: T^*X \rightarrow X$
 and identifies $KO^{-2}(pt)$ as the right place where the topological $\mathbb{Z}_2$ invariant really lives in.
   This fact matches perfectly with the results on the geometry of a Majorana zero mode, which is viewed as a $KR_{10}$-cycle with KO-dimension 2.
  In other words, the $\mathbb{Z}_2$ index is obtained from the pairing between an element in the $KR^{-2}$-group of the cotangent bundle
     and a reduced $KR_2$-cycle representing  a Majorana zero mode.
        \begin{equation}
     \xymatrixcolsep{5pc}\xymatrix{
  KR_2(T^*X) \times  KR^{-2}(T^*X) \ar@<4ex>[d]^{ind_t} \ar@<-5ex>[d]^{ind_a}  \ar[r]^-{index\,\, pairing}   & KO^{-2}(pt)  \ar[d]^{=}    \\
      KO^{-2}(pt) \times  KO^{-2}(pt) \ar[r]^-{f:\, \mathbb{Z}_2 \rightarrow \mathbb{Z}_2}  &    \mathbb{Z}_2     }
    \end{equation}
Here $f$ is a group homomorphism. There are two cases: (1) $f$ is the zero map, which maps $\{0, 1\}$ to $0$; (2) $f$ is the identity map, which maps $0$ to $0$ and $1$ to $1$. 
The map $f$ really represents the pairing between $ \mathbb{Z}_2 $  from homology and   cohomology. Since K-theory is dual to K-homology, the above index pairing
is essentially the Poincar{\'e} duality theorem. 

By the reduction from the effective Hamiltonian $\tilde{H}$ to the Dirac operator $D$, the geometry of a Majorana zero mode can be reduced to the $KR_5$-cycle of a chiral zero mode
$ (C^\infty(X),  \mathcal{L} \oplus  \Theta \mathcal{L} , D, \Theta )$, which determines a fundamental class  $[D, \Theta] \in KR_5(X)$. On the other hand, the time reversal gauge $w$ determines
a K-theoretic class  $[w] \in KQ^{-1}(X) = KR^{-5}(X)$. So the above index pairing can be reduced to a more familiar form as in the mod 2 index theorem. 
\begin{prop}
The index pairing for the topological $\mathbb{Z}_2$ invariant is  given by
   \begin{equation}
      \begin{array}{ccc}
       KR_5(X) \times KR^{-5}(X) & \rightarrow & KO^{-2}(pt) \\
       ([D, \Theta], [w]) & \mapsto &   \langle  [D] ,  [w] \rangle  
      \end{array}
   \end{equation}
where the right hand side can be spelled out as the mod 2 index theorem 
\begin{equation}
   sf(D, w^{-1}Dw) = ind_a(\tilde{H}) = ind_t(w) = \langle  [D] ,  [w] \rangle  
\end{equation}

\end{prop}

\section{Bulk-boundary correspondence}  
  In general, the bulk-boundary (or bulk-edge) correspondence is a realization
of the holographic principle in condensed matter physics. The philosophy of the
holographic principle, learned from the anti-de Sitter/conformal field theory (AdS/CFT) correspondence, is that the boundary field theory determines the geometry
in the bulk, and conversely the bulk geometry determines the boundary theory up
to some equivalence. Even though the general holographic principle still remains a
conjecture, the Chern--Simons/Wess--Zumino--Witten (CS/WZW) correspondence
is well understood as a theorem in mathematical physics.

The bulk-boundary correspondence can be understood on different levels. First of all, on the level of effective Hamiltonian, the boundary Hamiltonian can be obtained from
the  bulk Hamiltonian by a partial Fourier transform along a direction with
translational invariance. Such correspondence realized by the partial
Fourier transform holds for all dimensions, so there exists a one-to-one correspondence between the bulk and boundary Hamiltonians. 
In the topological band theory, we identify the bulk state with the corresponding boundary state via the partial Fourier transform. 
As a result, the geometry of this correspondence is trivial, since the partial Fourier transform always induces a map 
from the d-dimensional bulk to the $(d-1)$-dimensional boundary as momentum spaces. In terms of the topological $\mathbb{Z}_2$ invariant, if the bulk theory gives us $KO^{-2}(pt)$, 
then via this correspondence $KO^{-1}(pt)$ is obtained for the boundary of codimension $1$.
This is not the desired bulk-boundary correspondence, which is supposed to be an equivalence between effective (quantum) field theories from the bulk and boundary.

On the level of gauge theory, the Chern--Simons/Wess--Zumino--Witten
(CS/WZW) duality establishes an equivalence between the bulk and boundary field theories. 
For 3d topological insulators, the  $\mathbb{Z}_2$   invariant is the   WZW topological term in string theory
(or the odd topological index in index theory), which is
an important component of the Chern--Simons gauge theory. In other words, the bulk-boundary
correspondence on the level of action functional is given by the duality between
the bulk Chern--Simons action and the boundary   WZW term (also related to currrent algebra in vertex operator algebras). If we further
consider quantum field theories, then the bulk is described by a topological quantum
field theory (TQFT), and the boundary is characterized by a rational conformal
field theory (RCFT). So the bulk-boundary correspondence on the level of QFT/CFT
is given by CS3/WZW2 as a correspondence between the 3d CS TQFT and 2d WZW RCFT.
On the other hand, for a given gauge group $G$ such as $SU (2)$, there exists a
one-to-one correspondence between the WZW model and Chern--Simons theory,
both characterized by their level $k \in \mathbb{Z}$. If the relevant gauge theory is described by group cohomology,
then the bulk-boundary correspondence is given by the Dijkgraaf--Witten theory. If the
Dijkgraaf--Witten transgression map is $ \tau : H^4 (BG, \mathbb{Z}) \rightarrow  H^3 (G, \mathbb{Z})$,
then $\tau$ is the
correspondence between the Chern--Simons theory and WZW model associated with $G$. 

As different aspects of the topological $\mathbb{Z}_2$ invariant, the topological $\mathbb{Z}_2$ index  and the  Kane--Mele invariant are equivalent. 
The topological index  is an integral of the Chern character over the bulk, and the Kane--Mele invariant has a Pfaffian formalism derived from an effective fermionic field theory. 
In other words, the bulk theory is an action functional given by the WZW term, and the Kane--Mele invariant is an effective quantum field theory defined over the set of fixed points. 
We propose that the effective boundary is defined as the fixed points, which is different from the geometric boundary of the bulk. 
Hence the equivalence between the topological index and the Kane--Mele invariant is viewed as
the bulk-boundary correspondence on the level of topological invariant.

\begin{thm} 
   The  topological index $ind_t(w)$ and the Kane--Mele invariant $\nu$ are equivalent, 
   \begin{equation}
     \nu = (-1)^{ind_t(w)} 
   \end{equation}
   which is a  bulk-boundary correspondence if the set of fixed pionts $X^\tau$ is identified as the effective boundary.
\end{thm}

From the proof of the mod 2 index theorem, spectral flow is the key idea behind the odd topological index. 
For a Majorana zero mode, the chiral zero modes can pass through the zero energy level only at a fixed point. In the spirit of localization, 
the spectral flow of chiral states also indicates that the effective boundary is given by the fixed pionts. 
In a sense, the above equivalence is a fixed point theorem of the time reversal symmetry. 

Inspired by the Kane--Mele invariant defined over the fixed points, we propose   $X^\tau$ as the effective boundary.
The Kane--Mele invariant is a collective effect over $X^\tau$, so we further project it to an abstract point, i.e.,  $p: X^\tau \rightarrow pt$, as in the topological index map.  
Otherwise, we will obtain a family index theorem with values in $KO^{-2}(X^\tau)$ instead of a topological index in $KO^{-2}(pt)$. 
To stress the topological $\mathbb{Z}_2$ invariant $\nu$ as an index number, we realize the abstract point $\{ pt \}$ in the topological index map as a specific fixed point $x_0 \in X^\tau$
so that $\nu \in KO^{-2}(x_0)$.

\begin{examp}
    The $KR$-theory   of $\mathbb{T}^d$ can be computed  based on the decomposition of the fixed points
\begin{equation}
 (\mathbb{T}^d)^\tau = \oplus _{n =0}^d \binom{d}{n} \{ pt \}
\end{equation}
where $\binom{d}{k}$ is the binomial coefficient,  so that
\begin{equation}
 KR^{-j}(\mathbb{T}^d) = \oplus_{k =0}^d \binom{d}{k} KO^{k-j}(pt)
\end{equation}
The (strong) topological $\mathbb{Z}_2$ invariant only depends on the fixed point $x_0$ with ``top codimension'', i.e.,
\begin{equation}
   KO^{d-j}(x_0)  \quad \text{for} \,\,\, k =d
\end{equation}
In fact, one has the relation $d -j = -2$, for instance $j = 4, d= 2$,
\begin{equation}
   \nu \in \widetilde{KQ}(\mathbb{T}^2)  = \widetilde{KR}^{-4} (\mathbb{T}^2) = KO^{-2}(x_0)
\end{equation}

\end{examp}

In  the decomposition of the fixed points $(\mathbb{T}^d)^\tau$,
this specific fixed point $x_0 $ is taken as the fixed point corresponding to the last (i.e., $k= d$) summand, that is the fixed point of ``top codimension''.
This fixed point $x_0 \in X^\tau$ with ``top codimension'' is naturally identified with $\mathbb{S}^d$ $(d = \dim X)$ by the Poincar\'{e} duality in K-theory.

On the level of K-theory, the bulk-boundary correspondence is a KK-cycle connecting the bulk and boundary K-theories. 
The bulk is  the d-dimensional momentum space $X$, and   the effective boundary  is assumed to be the fixed points $X^\tau$ (or the fixed point with top codimension $x_0$). 
By the topological band theory, the bulk theory is modeled by the Quaternionic K-theory $KQ^{2-d}(X)$, which is canonically isomorphic to the Real K-theory $KR^{-2-d}(X)$.
Along the same logic, the boundary theory is modeled by the real K-theory $KO^{-2}(X^\tau)$ (or $KO^{-2}(x_0)$ as a number). Since the fixed points $X^\tau$ are real points 
with respect to the time reversal transformation $\tau$, one can identify $KR^{-2}(X^\tau) = KO^{-2}(X^\tau)$. 
Thus the bulk-boundary correspondence is a KKR-cycle connecting the KR-theory of the bulk and the KO-theory of the effective boundary,
\begin{equation}
     \xymatrixcolsep{1.5pc}\xymatrix{
      KR^{-2-d}(X) \times KKR^{d}(X, X^\tau) \ar[r] & KO^{-2}(X^\tau) \ar[r]  & KO^{-2}(x_0)   }  
\end{equation}
which is a generalization of the topological index map
\begin{equation}
   ind_t :  KR^{-2-d}(X) \rightarrow KO^{-2}(pt)
\end{equation}

\begin{examp} 
    When $X = \mathbb{T}^d$ is a torus, it is identified as $\mathbb{T}^d = \mathbb{R}^d /\mathbb{Z}^d$ by the translational symmetry group $\Gamma = \mathbb{Z}^d$. 
    The real assembly map $ \mu_\mathbb{R}$ in the real  Baum--Connes conjecture for $\Gamma = \mathbb{Z}^d$ is an isomorphism, called the Baum--Connes isomorphism or dual-Dirac isomorphism,  
       \begin{equation}
      \mu_\mathbb{R}(\mathbb{Z}^d):  KO_i(\mathbb{T}^d)   \cong KO_i(C^*(\mathbb{Z}^d, \mathbb{R})) 
   \end{equation}
   On the right hand side, the KO-theory of the group $C^*$-algebra $C^*(\mathbb{Z}^d, \mathbb{R})$ is
   \begin{equation}
           KR^{-i} (\mathbb{T}^d, \tau) =  KO_i(C(\mathbb{T}^d, \tau))  =  KO_i(C^*(\mathbb{Z}^d, \mathbb{R})) 
   \end{equation}
   since the KR-theory of $(\mathbb{T}^d, \tau)$ can be identified with the  KO-theory of the real function algebra  $C(\mathbb{T}^d, \tau)$, where
   \begin{equation}
    C(\mathbb{T}^d, \tau) = \{f \in C (\mathbb{T}^d ) \, |\,  \overline{f (x)} = f (\tau (x))\}
   \end{equation}
In fact, the dual-Dirac isomorphism can be realized by an invertible real KKO-class $\beta \in KKO(C_\mathbb{R}(\mathbb{T}^d), C(\mathbb{T}^d, \tau))$,
\begin{equation}
 KO_i(\mathbb{T}^d) \times KKO(C_\mathbb{R}(\mathbb{T}^d), C(\mathbb{T}^d, \tau)) \xrightarrow{\simeq} KO_i(C(\mathbb{T}^d, \tau))
\end{equation}
Let $\alpha$ be the inverse  of $\beta$,
that is, $\alpha \in KKO( C(\mathbb{T}^d, \tau), C_\mathbb{R}(\mathbb{T}^d)) $ such that $\alpha \circ \beta = id$,
so $\alpha$ realizes the Dirac isomorphism
\begin{equation}
 KR^{-i} (\mathbb{T}^d, \tau) \times KKO( C(\mathbb{T}^d, \tau), C_\mathbb{R}(\mathbb{T}^d)) \xrightarrow{\simeq} KO_i(\mathbb{T}^d) %\cong KO^{d-i}(\mathbb{T}^d)
\end{equation}
In addition, if $i_0: x_0 \hookrightarrow \mathbb{T}^d$ is the inclusion map, then it induces the restriction map in $KO$-theory
\begin{equation}
 i_0^*:  KO^{-j}(\mathbb{T}^d) \rightarrow KO^{-j} (x_0)
\end{equation}

  In this setting,   the bulk-boundary correspondence for the torus $\mathbb{T}^d$  is a surjective map realized as the composition map  
   \begin{equation}
     i_0^* \circ PD \circ \alpha: \,\,  \widetilde{KR}^{-i} (\mathbb{T}^d) \stackrel{\alpha}{\rightarrow} \widetilde{KO}_i(\mathbb{T}^d) \stackrel{PD}{\rightarrow}  \widetilde{KO}^{d-i}(\mathbb{T}^d) \stackrel{i_0^*}{\rightarrow}   KO^{-2}(pt)
   \end{equation}
   where 
   the middle isomorphism is the   Poincar\'{e} duality in $KO$-theory.
      In this case, the Dirac isomorphism is applied to convert the KR-theory of the bulk to a KO-homology, then the Poincar{\'e} duality is used to get back the KO-theory of the bulk.
   So the KKR-cycle is reduced to a KKO-cycle (via isomorphisms) to realize the bulk-boundary correspondence between the KO-theories of the bulk and boundary.
\end{examp}

From the above, the bulk-boundary correspondence on the level of K-theory is essentially a KKO-cycle or KKR-cycle connecting the bulk and boundary K-theories, which is not necessary
to be an isomorphism. 

On the other hand, we  have seen that it is interesting to understand the bulk theory by the geometry of Majorana zero modes using $KR$-cycles or $KR$-homology,
a similar idea is carried out in \cite{BCR16}. So the bulk-boundary
correspondence in $KR$-homology  has the following form,
\begin{equation}
   KKR(X^\tau, X) \times KR_2(X) \rightarrow KO_{2}(X^\tau) \xrightarrow{\simeq}   KO^{-2}(X^\tau) {\rightarrow} KO^{-2}(x_0)
\end{equation}
where the isomorphism is again the Poincar\'{e} duality. This bulk-boundary correspondence on the level of K-homology is viewed as a generalized version of the
analytical index of the $KR_2$-cycle of the effective Hamiltonian.

Therefore, if the bulk-boundary correspondence is realized as a correspondence connecting K-homologies, then it is a generalization of the analytical index map;
if the bulk-boundary correspondence is realized as a correspondence connecting K-theories, then it is a generalization of the topological index map.
Finally, we have the following diagram based on Kasparov products if  K-homology and K-theory are written in $KKR$-theory or $KKO$-theory,
\begin{equation}
   \xymatrixcolsep{5pc}\xymatrix{
  KKR(X^\tau, X) \times KKR_2(X, \mathbb{C} )   \ar[r]   & KKO_{2}(X^\tau, \mathbb{R})  \ar[d]^{\cong}    \\
      KKR^{-d-2}(\mathbb{C}, X) \times KKR^{d}(X, X^\tau) \ar[r]  &   KKO^{-2}(\mathbb{R}, X^\tau)   }
\end{equation}

\begin{examp}
In particular, the bulk-boundary correspondence offers a new perspective to view the
mod 2 index theorem behind the topological $\mathbb{Z}_2$ invariant.
 \begin{equation}
   \xymatrixcolsep{5pc}\xymatrix{
  KKR(X^\tau, X) \times KR_2(X)   \ar[r]^-{ind_a}   & KO^{-2}(x_0)  \ar[d]^{=}    \\
      KQ^{2-d}(X) \times KKR^{d}(X, X^\tau) \ar[r]^-{ind_t}  &   KO^{-2}(x_0)   }
\end{equation}
\end{examp}

\section{Family index theorem}
Let us briefly review the relation between the determinant and  Pfaffian line bundles and a family index theorem, and apply it to the index problem of Majorana zero modes. 
The classical results on the determinant line bundle in the context of index theory can be found in \cite{F87}. This section gives a preparation for the proof of the equivalence between
the noncommutative topological index and the noncommutative Kane--Mele invariant.

Let $Y$ be a compact spin manifold, and $D_+: \mathcal{H}_+ \rightarrow \mathcal{H}_- $ be a Dirac operator mapping $\mathcal{H}_+$ to $\mathcal{H}_-$,
where $\mathcal{H}_+$ and $\mathcal{H}_-$ are finite dimensional Hilbert spaces of spinors. Taking the top exterior products of the Hilbert spaces, denoted by
$\det \mathcal{H}_\pm = \wedge^{max} \mathcal{H}_\pm$, $D_+$ induces a natural map
\begin{equation}
 \det D_+ : \det \mathcal{H}_+ \rightarrow \det \mathcal{H}_-
\end{equation}
Or equivalently, $\det D_+$ can be viewed as an endomorphism
\begin{equation}
 \det D_+ \in  (\det \mathcal{H}_+)^* \otimes \det \mathcal{H}_-
\end{equation}

The above construction can be promoted to vector bundles. Let $\pi: Z \rightarrow X$ be a vector bundle 
with a typical fiber $Z_x \cong Y$, which is assumed to
be spin. Now let $D_+$ be a family of  Fredholm operators parametrized by the base manifold $X$, 
\begin{equation}
 D_{+,x} :  (\mathcal{H}_+)_x \rightarrow (\mathcal{H}_-)_x 
\end{equation} 
In other words, $D_+$ is a bundle map between two Hilbert bundles,
 \begin{equation}
  \xymatrix{
  \mathcal{H}_+ \ar[rd]  \ar@{-}[rr]^{D_+} & \ar[r] & \mathcal{H}_-  \ar[ld] \\
          &  X  & }
 \end{equation}
 Define the determinant line bundle of $D_+$ as a complex line bundle,
 \begin{equation}
   \pi: Det D_+ \rightarrow X
 \end{equation}
 such that each fiber takes the form
 \begin{equation}
   (DetD_+)_x = (\det \ker D_{+,x})^* \otimes \det \ker D_{+,x}, \quad x \in X
 \end{equation}
 where the Hilbert spaces have been identified as
 \begin{equation}
     (\mathcal{H}_+)_x \cong  (\ker D_{+,x})^*, \quad  (\mathcal{H}_-)_x \cong \ker D_{+,x}
 \end{equation}

On the other hand, for each $x \in X$, the Fredholm operator $D_{+,x}$ has finite dimensional kernel and cokernel, and 
the Fredholm index is the analytical index,
\begin{equation}
 index(D_{+,x}) = \dim ker \, D_{+,x} - \dim ker \, D^*_{+,x}
\end{equation}
where $ D^*_{+,x}$ is the adjoint of  $D_{+,x}$.
Therefore, one defines the index bundle as the formal difference bundle,
\begin{equation}
 \pi: Ind(D_+) \rightarrow X, \quad Ind (D_+) = ker D_+ \ominus ker D^*_+
\end{equation}
where $kerD_+$ and $ker D^*_+$ are vector bundles over $X$. 
Obviously, the index bundle of $D_+$ gives rise to an element in the K-theory of the base manifold,
\begin{equation}
 [Ind(D_+)] \in K(X)
\end{equation}

Putting it together, the determinant line bundle of the index bundle $\pi: Ind(D_+) \rightarrow X$ is isomorphic 
to the determinant line bundle of the family of Fredholm operators $D_+$,
\begin{equation}
 det(Ind(D_+)) \cong Det(D_+)
\end{equation}
As a consequence,
 the first Chern class of the index bundle $Ind(D_+)$ can be represented by the determinant line bundle of $D_+$,
\begin{equation}
 c_1(Ind(D_+)) = [DetD_+] \in H^2(X, \mathbb{Z})
\end{equation}

If the typical fiber $Y$ is spin and $(8k +2)$-dimensional, then there exists a Pfaffian line bundle, 
\begin{equation}
 \pi: Pf(D_+) \rightarrow X
\end{equation}
such that 
\begin{equation}
 Pf(D_+) \otimes Pf(D_+) \cong Det(D_+)
\end{equation}
There exists a push-forward map,  called the Gysin map, between the real K-groups of the total and base spaces in $\pi: Z \rightarrow X$,
\begin{equation}
 \pi_!: KO(Z) \rightarrow KO^{-2}(X)
\end{equation}
If one considers the index bundle $\pi: Ind(D_+) \rightarrow X$ in this real case, then one has the identity,
\begin{equation}
  \pi_!([1]) = [Ind(D_+)] \in KO^{-2}(X)
\end{equation}
where $[1]$ is a certain generator in $KO(Z)$ \cite{F87}. 

Moreover, by the Chern--Weil theory, there exists a map induced by the Pfaffian form, 
\begin{equation}
 Pfaff: KO^{-2}(X) \rightarrow H^2(X, \mathbb{Z})
\end{equation}
where $H^2(X, \mathbb{Z})$ is the 2nd cohomology group characterizing complex line bundles over $X$.
By a theorem in \cite{F87}, one has the relation,
\begin{equation}
 c_1(Pf(D_+)) = Pfaff([Ind(D_+)]) \in H^2(X, \mathbb{Z})
\end{equation}

Now we apply the above construction to the effective Hamiltonian of Majorana zero modes,
which is a family of skew-adjoint Fredholm operators, 
\begin{equation}
 \tilde{D}(x) = \begin{pmatrix}
                                                 0 &  - D(x)  \\
                                                 D(x) & 0
                                                \end{pmatrix}
\end{equation}
 We define the determinant line bundle of $\tilde{D}$ as
\begin{equation}
   Det(\tilde{D}) =  (\det ker \, D )^* \otimes \det ker \, (-D) 
\end{equation}
Suppose $ker D= ker (-D)$ is one dimensional, in this case, 
\begin{equation}
 Det(\tilde{D})  \cong  Det(\mathcal{H})
\end{equation}
By the decomposition of the skew-adjoint operator $\tilde{D}$,
there exists a Pfaffian line bundle $\pi: Pf(\tilde{D}) \rightarrow X$, which is a generalization of the Pfaffian line bundle $\pi: Pf(\mathcal{H}) \rightarrow X$ in the previous subsection.
%Since the first Chern class of the Hilbert bundle $\pi: \mathcal{H} \rightarrow X$ is a 2-torsion, i.e., $2c_1(\mathcal{H}) = 0$,

In addition, $D$ and $-D  $ have zero modes only at the fixed points $X^\tau$, 
the determinant line bundle $Det(\tilde{D})$ is restricted to a vector bundle over $X^\tau$,
\begin{equation}
 \pi: Det(\tilde{D})|_{X^\tau} \rightarrow X^\tau
\end{equation}
By definition, the Pfaffian line bundle is then a real line bundle,
\begin{equation}
 \pi: Pf(\tilde{D}) \rightarrow X^\tau
\end{equation}
such that 
\begin{equation}
 Pf(\tilde{D}) \otimes Pf(\tilde{D}) \cong Det(\tilde{D})|_{X^\tau}
\end{equation}
Therefore, the skew-adjoint operator $\tilde{D}$ has the family index, 
\begin{equation}
 [Ind(\tilde{D})] \in KO^{-2}(X^\tau) 
\end{equation}

%and the first Chern class is %reduced to the Stiefel--Whitney class \cite{KLW16},
%$$
%c_1(Pf(\tilde{D})) = Pfaff([Ind(\tilde{D})]) \in H^2(X^\tau, \mathbb{Z}_2) %\rightarrow H^2(X^\tau, \mathbb{Z}_2) \ni sw_2(Pf(D))
%$$

In order to get the topological index (as  a number) from the above family index, one has to project the set of fixed points to an abstract point $p: X^\tau \rightarrow pt$,
which induces the push-forward map in real K-theory $p_*: KO^{-2}(X^\tau) \rightarrow KO^{-2}(pt)$. As mentioned above, we realized the abstract point $\{pt\}$
as a specific fixed point $x_0 \in X^\tau$ with top codimension, %which is viewed as the effective boundary in geometry. And in this case, 
and the projection $p_*$ can be
viewed as a summation $\sum: KO^{-2}(X^\tau) \rightarrow KO^{-2}(x_0)$ to count the collective effect over the fixed points.

\section{Noncommutative topology index}
 
  As a modern geometric framework, noncommutative geometry (NCG) has been providing computational tools for studying topological invariants in disordered fermionic systems, 
   e.g. the noncommutative model of quantum Hall effect \cite{BES94} and disordered topological insulators \cite{P11, S16}.
  We generalize the index theory and K-theory of the topological $\mathbb{Z}_2$ invariant onto the noncommutative 2-torus $\mathbb{T}^2_\theta$ and 3-sphere $S^3_\theta$ \cite{KLW1601}.

When the base manifold  is generalized to be a noncommutative manifold defined by a $C^*$-algebra,
the topological index can be defined with the help of the Connes--Chern characters. 
As a generalization of the Chern character from K-theory to de-Rham cohomology, the Connes--Chern characters are maps from  K-theory (or K-homology) to cyclic homology (or cyclic cohomology). 
By the machinery of noncommutative topology, the topological index can be computed
by pairing periodic cyclic cohomology and homology after applying the Connes--Chern characters \cite{C94, K09}. 
In this section, we define the noncommutative $\mathbb{Z}_2$ invariant for disordered topological insulators
as a topological index in noncommutative topology.

 In order to fix the notations, let us briefly review  the Connes--Chern characters acting on K-theory and K-homology, 
 and we only care about the odd case motivated by the mod 2 index theorem behind the topological $\mathbb{Z}_2$ invariant. 
 
 Let $A$ be a unital $C^*$-algebra representing a noncommutative manifold,
  the odd Connes--Chern character map from  K-theory to cyclic homology groups is defined by
 \begin{equation}
   \quad  Ch_{2n+1} : K_1(A) \rightarrow HC_{2n+1}(A), \quad [u] \mapsto Tr(u^{-1}\otimes u \otimes \cdots\otimes u^{-1} \otimes u)
 \end{equation}
  where $Tr$ is the operator trace, and there are $(n+1)$-pairs of unitary operators $u$ and $u^{-1}$ in the trace.
 It is convenient to equivalently define the above Connes--Chern character by a pairing between cyclic cohomology (dual to cyclic homology) and K-theory,
 \begin{equation}
    HC^{2n+1}(A) \times K_1(A) \rightarrow \mathbb{C}
 \end{equation}
 Furthermore,  it induces a pairing between periodic cyclic cohomology with K-theory,
 \begin{equation}
    HP^1(A) \times  K_1(A)   \rightarrow \mathbb{C}
 \end{equation}
 More precisely, for an odd $(b, B)$-cocycle $\phi  = (\phi_1, \phi_3, \cdots, \phi_{2k+1}, \cdots )$ 
 and a unitary element $u \in A$, the pairing between $HP^1(A)$ and  $K_1(A)$  is defined by 
 \begin{equation}
   \langle [\phi], [u] \rangle = \frac{1}{\Gamma(\frac{1}{2})}\sum_{k = 0}^\infty {(-1)^{k+1}} k! \phi_{2k+1}(u^{-1}, u, \cdots, u^{-1}, u)
 \end{equation}

 Let $(\mathcal{H}, F)$  be an odd $p$-summable Fredholm module over a $C^*$-algebra $A$ and let $n$ be an integer such that $2n \geq p$.
 The odd Connes--Chern character of $(\mathcal{H},F)$, denoted by $Ch^{2m-1}(\mathcal{H}, F)$, is defined by 
 \begin{equation}
   Ch^{2m-1}(\mathcal{H}, F) (a_0, \cdots, a_{2m-1}) = (-1)^m \frac{2 \Gamma(m - \frac{1}{2})}{\Gamma(\frac{1}{2})} Tr(F[F, a_0] \cdots [F, a_{2m-1}])
 \end{equation}
  For any integer $m \geq n$, the class of the cyclic cocycle $Ch^{2m-1}(\mathcal{H}, F)$ is stabilized in the odd periodic 
 cyclic cohomology group, call it the stable odd Connes--Chern character of $(\mathcal{H},F)$, i.e., 
 \begin{equation}
   Ch^{1} (\mathcal{H}, F) = [Ch^{2m-1}(\mathcal{H}, F)] \in HP^1(A), \quad m \geq n
 \end{equation}
 In other words, the Connes--Chern character defines a map from K-homology to periodic cyclic cohomology,
 \begin{equation}
     Ch^{1} : K^1(A) \rightarrow HP^1(A), \quad [(\mathcal{H},F)] \mapsto Ch^{1} (\mathcal{H}, F)
 \end{equation}

 As a remark, given a spectral triple $(A, \mathcal{H}, D)$, if the Dirac operator $D$ is invertible, then one defines $F = D|D|^{-1}$ as the phase of $D$,
 and the triple $(A, \mathcal{H}, F)$ is a Fredholm module.

Let $(\mathcal{H}, F)$ be an odd  Fredholm module over $A$ and let $u \in A$ be an invertible element. Define  
the (positive) projection operator by,
\begin{equation}
 P = \frac{1+ F}{2}: \mathcal{H} \rightarrow \mathcal{H}
\end{equation}
so that $PuP: P\mathcal{H} \rightarrow P\mathcal{H}$ is a Fredholm operator.
Of course, it is also possible to define the (negative) projection by $P = (1-F)/2$.
Now the Fredholm index of $PuP$ can be computed by 
\begin{equation}
 index(PuP) = \frac{(-1)^n}{2^{n}} Tr(F[F, u^{-1}][F, u] \cdots [F, u^{-1}][F, u])
\end{equation}
Alternatively, using $du= [F, u]$ and $du^{-1} = [F, u^{-1}]$, 
\begin{equation}
 index(PuP) = \frac{(-1)^n}{2^{n}} Tr[F(du^{-1}du)^n]
\end{equation}

\begin{examp}
   Let $(A, \mathcal{H}, D)$ be a 3-summable spectral triple and $(A, \mathcal{H}, F)$ be the associated Fredholm module,  one defines 
 the projection $P = (1+F)/2$ as usual. 
 For a unitary  operator $u \in U(A)$ (or $[u] \in K_1(A)$), the Fredholm index of $PuP$ can be computed by 
 \begin{equation}
   ind(PuP) = - \frac{1}{8}Tr[F( du^{-1}  du)^3] = -\frac{1}{4}Tr[( u^{-1}  du)^3]
 \end{equation}
 with $da = [F, a]$ for $a \in A$.
 \end{examp}
 
Furthermore, using the pairing between cyclic cohomology and  homology,
 \begin{equation}
   \langle \,\, , \,\, \rangle: HC^{2n-1}(A) \times HC_{2n-1} (A) \rightarrow \mathbb{C}
 \end{equation}
 the above Fredholm index can be computed by the pairing,
 \begin{equation}
    index(PuP) = \langle Ch^{2n-1}(H, F), Ch_{2n-1}(u) \rangle 
 \end{equation}
which is further written as
\begin{equation}
    index(PuP) = \langle Ch^{1}(H, F), Ch_{1}(u) \rangle 
\end{equation}
provided $n$ is in the stable range (i.e., $2n \geq p$).

The above discussion  can be  summarized in the commutative diagram \cite{K09},
 \begin{equation} \label{CommDiag}
  \xymatrixcolsep{5pc}\xymatrix{
  K^1(A) \times  K_1(A) \ar@<4ex>[d]^{Ch_1} \ar@<-5ex>[d]^{Ch^1}  \ar[r]^-{index\,\, pairing}   & \mathbb{Z}  \ar[d]    \\
      HP^1(A) \times  HP_1(A) \ar[r]^-{pairing}  &  \mathbb{C}     }
 \end{equation}

Now we  apply the machinery of noncommutative topology to define the noncommutative  $\mathbb{Z}_2$ invariant as a topological index. To proceed,  we first model Majorana zero modes by a KQ-cycle over $C^*$-algebras and  then use the relevant data  to define a spectral triple.  
 Then the topological $\mathbb{Z}_2$ index will be defined based on the pairing between periodic cyclic homology and cohomology  after
 applying the Connes--Chern characters. 
 
 Let $A$ be a  $C^*$-algebra  and $\mathcal{H}$ be a  Hilbert space, 
 suppose there is a representation $\pi: A \rightarrow B(\mathcal{H})$, so that $A$ is a noncommutative space of bounded operators on $\mathcal{H}$.
 Time reversal symmetry introduces the time reversal operator $\Theta$ acting on both $\mathcal{H}$ and $A$, and $\Theta$ is assumed to be an anti-unitary operator which satisfies $\Theta^2 = -1$.
 With the time reversal operator, $(\mathcal{H}, \Theta)$ can be viewed as a Hilbert space over quaternions $\mathbb{H}$, denoted by $\mathcal{H}_\mathbb{H} = (\mathcal{H}, \Theta)$.
 Or equivalently, $\mathcal{H}_\mathbb{H}$ can be decomposed as $\mathcal{H}_\mathbb{H} = \mathcal{H}_\mathbb{C} \oplus \Theta \mathcal{H}_\mathbb{C} $.
 The pair $(A, \Theta)$ is called a \emph{Real} $C^*$-algebra with the  real structure  
 $\Theta$, and a topological insulator can be described by $(A, \Theta)$ \cite{K15}.
  One further  considers the  KQ-theory of $(A, \Theta)$, denoted by $KQ_*(A) = KQ_*(A, \Theta)$, which can in turn be computed by KR-theory  $KR_*(A)$ \cite{R15}. 
 
 \begin{examp}
   Let $\theta \in \mathbb{R} \setminus \mathbb{Q} $ be an irrational real parameter,
     the noncommutative 2-torus $\mathbb{T}_\theta^2 = C^*(u,v)$ is the universal $C^*$-algebra generated by two unitary operators
 $u,v$  satisfying $uv = e^{2 \pi i \theta} vu$.
 One concrete realization of  $\mathbb{T}_\theta^2$ is given by the irrational rotation algebra or $C^*$-crossed product $C(S^1) \rtimes \mathbb{Z}$.  
 Let  $ L^2(S^1)$ be the Hilbert space, the irrational rotation algebra is generated by $C(S^1)$ and the rotation action by $\theta$. 
 More precisely, for any $g \in L^2(S^1) $, two unitary generators are defined by $U = M_f$ for $f \in C(S^1)$ and $V = V_\theta$,
 \begin{equation}
   M_fg = fg, \quad V_\theta g(z) = g(e^{-2\pi i \theta} z)
 \end{equation}

 \end{examp}

 \begin{examp}
    The noncommutative 3-sphere $S^3_\theta$ is defined as the universal $C^*$-algebra generated by two normal
operators $\alpha$ and $\beta$ (s.t. $\alpha \alpha^* =  \alpha^* \alpha$,  $\beta \beta^* = \beta^* \beta$) satisfying the relations,
\begin{equation}
 \alpha \beta = \lambda \beta \alpha, \quad  \alpha^* \beta = \bar{\lambda} \beta \alpha^*,  \quad \alpha \alpha^* + \beta \beta^* = 1
\end{equation}
for the complex parameter $\lambda = e^{2\pi i \theta}$ and irrational $\theta \in \mathbb{R} \setminus \mathbb{Q}$.
There exists a natural parametrization of the generators in $S^3_\theta$ by Hopf coordinates,
\begin{equation}
 \alpha = u \cos \psi , \quad \beta = v  \sin \psi, \quad \psi \in [0, \pi/2]
\end{equation}
where $u,v$ are the generators of the noncommutative 2-torus $\mathbb{T}_\theta^2$.
 \end{examp}

 \begin{defn}
    The time reversal symmetry is defined as an anti-linear automorphism $\Theta$  on the given $C^*$-algebra $A$ such that  $\Theta^2 = -1$, so $\Theta$ is a  real structure and $(A, \Theta)$ is a Real $C^*$-algebra.
 \end{defn}

 \begin{examp} 
 If the $C^*$-algebra $A$ is generated by two operators, then 
 the time reversal symmetry  is defined by $\Theta := i\sigma_2* = \begin{pmatrix}
                 0 & * \\
                 -* & 0
              \end{pmatrix}$,
   where $*(a) = a^*$ is the adjoint operation.
 \end{examp}

 Let $\tilde{D}$ be a localized skew-adjoint Fredholm operator  approximating the effective Hamiltonian of a topological insulator,
 \begin{equation}
  \tilde{D} := \begin{pmatrix}
       0 & -D \\
       D & 0
     \end{pmatrix} : \,\, \begin{matrix}
                      \mathcal{H}_\mathbb{C} \\
                         \oplus \\
                         \Theta \mathcal{H}_\mathbb{C}
                         \end{matrix} \rightarrow 
                             \begin{matrix}
                            \Theta  \mathcal{H}_\mathbb{C} \\
                             \oplus \\
                            \mathcal{H}_\mathbb{C}
                             \end{matrix}
 \end{equation}
 where $D$ is  a self-adjoint Dirac operator acting on $\mathcal{H}_\mathbb{C}$.
 $\tilde{D}$ is an unbounded operator if and only if $D$ is an unbounded operator. 
$\tilde{D}$ is a skew-adjoint (i.e., $\tilde{D}^* = - \tilde{D}$) Fredholm operator, so its analytical index is  
 \begin{equation}\label{analind}
    ind_a(\tilde{D})   = \dim_\mathbb{H} \ker \tilde{D} \equiv \dim_\mathbb{C} \ker D \quad \text{(mod 2)} %\in KO^{-1}(pt)
 \end{equation}

As before,  we define the real structure   
 $  \mathcal{J} = \begin{pmatrix}
         0 & \Theta^* \\
         \Theta & 0
      \end{pmatrix} $ which statisfies $ \mathcal{J} =  \mathcal{J}^*$ and $ \mathcal{J}^2 = 1$, and the grading operator 
      $\gamma = \begin{pmatrix} 
                  1 & 0  \\
                  0 & -1
                \end{pmatrix} $ as usual.
 The quintuple $(A, \mathcal{H} \oplus \Theta \mathcal{H}, \tilde{D},  \mathcal{J}, \gamma )$, called the $KQ$-cycle of Majorana zero modes, 
  has an associated 
   KR-cycle of  chiral zero modes $(A, \mathcal{H}, D, \Theta)$.

 \begin{defn}
  For a 3-summable spectral triple $(A, \mathcal{H}, D)$ associated to the $KQ$-cycle of Majorana zero modes
 $(A, \mathcal{H} \oplus \Theta \mathcal{H} , \tilde{D}, J, \gamma)$ and an element  $[w] \in KQ_1(A)$ that generates a $\mathbb{Z}_2$ component,
 the 3d noncommutative topological $\mathbb{Z}_2$ invariant is defined as the odd topological index,
  \begin{equation} \label{3dTopInd}
   ind(PwP) = \langle Ch^1(D), Ch_1(w) \rangle = \frac{1}{4 \pi^2}Tr[( w^{-1} [F, w])^3] %\quad \text{(mod 2)}
 \end{equation}
where the phase of $D$ is $F = D|D|^{-1}$ and the projection $P = (1-F)/2$.
 
 \end{defn}

For a topological insulator, we are interested in the occupied bands below the Fermi energy, which is assumed to be located at the zero energy level, we define 
the (Fermi) projection $P$ as the negative projection.
By assumption  $[w] \in KQ_1(A)$ is a  $\mathbb{Z}_2$ generator in the odd KQ-group, which generalizes the transition function of the Hilbert bundle.
If $A = C^\infty(\mathbb{T}^3)$ is the smooth function algebra on the 3-torus, 
$ind(PwP)$ gives us the strong $\mathbb{Z}_2$ invariant as in the classical case.

 Hence the Fredholm index of $PwP$ 
is computed by the index pairing between  $[(\mathcal{H}, D)]$ (KR-homology class of the $KR_5$-cycle $(A, \mathcal{H}, D, \Theta)$) and  $[w] \in KR_5(A) = KQ_1(A)$,
\begin{equation}
   ind(PwP) = \langle [(\mathcal{H}, D)], [w] \rangle \in KO_1(\mathbb{R}) \cong \mathbb{Z}_2
\end{equation}
which is intrinsically $\mathbb{Z}_2$-valued.

In the classical case, the even Chern character modulo 2 cannot give the topological $\mathbb{Z}_2$ invariant for 2d spaces such as $\mathbb{T}^2$.
By the free  $\mathbb{Z}_2$-action, $\mathbb{T}^2$ can be decomposed into two cylinders, and the Hilbert bundle is determined by the clutching function $w: S^1 \sqcup S^1 \rightarrow U(2)$, 
where $S^1 \sqcup S^1 = S^1 \times \mathbb{Z}_2$ is the boundary of a cylinder $S^1 \times I$. The topological index is a resolution of the gauge anomaly as a boundary term, see Prop. \ref{2dtopindex}, 
\begin{equation}
 ind(w) = \frac{1}{2\pi} \int_{S^1 \times \mathbb{Z}_2} tr(w^{-1}dw)
\end{equation}

The  topological $\mathbb{Z}_2$ index on the noncommutative 2-torus $\mathbb{T}^2_\theta$ can be defined similarly. 
Consider the irrational rotation algebra $C(S^1) \rtimes \mathbb{Z} = C^*(U, V) = C^*(M_f, V_\theta)$ for any non-zero continuous function $f \in C(S^1)$, the action of the time reversal operator $\Theta$ is spelled out as,
\begin{equation}
   \Theta U  = V^*_\theta = V_{-\theta}, \quad \Theta V = -U^* = -M_f^* = M_{-f^{-1}}
\end{equation}
Due to the effect $\Theta: \theta \mapsto - \theta$,  $C(S^1) \rtimes \mathbb{Z}$ can be decomposed into two copies of  $C(S^1) \rtimes \mathbb{N}_0$ ($\mathbb{N}_0 = \mathbb{N} \cup \{ 0 \}$) analogous to the decomposition of $\mathbb{T}^2$. 
Of course the fixed points of $\Theta$ on the real parameter are $\theta = \{ 0, \infty \}$, that is, the boundary of $C(S^1) \rtimes \mathbb{N}_0$ is $ C(S^1) \times \{ 0, \infty \} = C(S^1) \times \mathbb{Z}_2$. 
Let $w$ be a unitary operator on $C(S^1) \times \mathbb{Z}_2$ representing the time reversal operator $\Theta$, i.e., $w = \Theta K$ ($K$ is the complex conjugation), 
then the topological $\mathbb{Z}_2$ index can be defined by dimensional reduction from $\mathbb{T}^2_\theta$ to $C(S^1) \times \mathbb{Z}_2$, which gives a generalization of the classical case. 

\begin{defn}
   For a 2-summable spectral triple $(C^\infty(\mathbb{T}^2_\theta), \mathcal{H}, D)$ associated to the $KQ$-cycle of Majorana zero modes
 $(C^\infty(\mathbb{T}^2_\theta), \mathcal{H} \oplus \Theta \mathcal{H} , \tilde{D}, \mathcal{J}, \gamma)$ on the noncommutative 2-torus $\mathbb{T}^2_\theta$,
 the boundary Dirac operator, denoted by $D_\partial$, is defined by restriction  $ D_\partial := D|_{C(S^1) \times \mathbb{Z}_2} $.
 If  $[w] \in KQ_1(C(S^1) \times \mathbb{Z}_2)$ generates a $\mathbb{Z}_2$ element, then
 the 2d noncommutative topological $\mathbb{Z}_2$ invariant is defined by pairing boundary classes,
  \begin{equation} \label{2dNCTorus}
   ind(P_\partial wP_\partial) = \langle Ch^1(D_\partial), Ch_1(w) \rangle = \frac{1}{2 \pi}Tr( w^{-1} [F_\partial, w]) %\quad \text{(mod 2)}
 \end{equation}
where the phase of $D_\partial$ is $F_\partial = D_\partial|D_\partial|^{-1}$ and the projection $P_\partial = (1-F_\partial)/2$.
\end{defn}

\section{Fixed point algebra}
 
We have seen the noncommutative 2-torus $\mathbb{T}^2_\theta$ and 3-sphere $S^3_\theta$ as concrete examples, 
the time reversal operator has been introduced for such noncommutative manifolds with two generators.
We will define the fixed point $C^*$-algebra of the time reversal symmetry, which is the noncommutative analog of the set of fixed points. 
We mainly focus on  $\mathbb{T}^2_\theta$, and the fixed point algebra in $S^3_\theta$ can be treated similarly.

In the classical case, if the 2-torus $\mathbb{T}^2$ is parametrized by the angles
   \begin{equation}
       \mathbb{T}^2 = \{ (e^{i\theta_1}, e^{i\theta_2}) ~|~ \theta_i \in (-\pi, \pi ], i = 1, 2 \}
   \end{equation}
   then the time reversal transformation on $\mathbb{T}^2$ is defined by 
   \begin{equation}
       \tau:  \mathbb{T}^2 \rightarrow \mathbb{T}^2; \quad (e^{i\theta_1}, e^{i\theta_2}) \mapsto (e^{-i\theta_1}, e^{-i\theta_2})
   \end{equation}
So $(\mathbb{T}^2, \tau)$ is a Real space with the real structure $\tau$ (s.t. $\tau^2 = 1$), and the set of fixed points is $\mathbb{T}^\tau = \{(0, 0), (0, \pi), (\pi, 0), (\pi, \pi) \}$.
Let $C(\mathbb{T}^2)$ denote the continuous functions on $\mathbb{T}^2$, if the complex conjugation is denoted by $K$,
\begin{equation}
 K: C(\mathbb{T}^2) \rightarrow C(\mathbb{T}^2), \quad K(f) = \bar{f}
\end{equation}
then $(C(\mathbb{T}^2), K)$ is a Real function algebra with the real structure $K$ (s.t. $K^2 = 1$). The invariant subspace of $C(\mathbb{T}^2)$ with respect to $\tau$ and $K$ is defined as
\begin{equation} 
   C_{R} (\mathbb{T}^2):= \{ f\in C(\mathbb{T}^2) ~|~ K \circ f = f \circ \tau \,\, or\,\, \overline{f(x)} = f(\tau(x))  \}
\end{equation}
which is a  commutative real  $C^*$-algebra \cite{R15}.

  The time reversal operator $\Theta$ is  an anti-unitary operator acting on a $C^*$-algebra such that $\Theta^2 = -1$.
  $\Theta$ acting on $\mathbb{T}_\theta^2$ can be  defined as
 \begin{equation}
    \Theta := \begin{pmatrix}
                 0 & * \\
                 -* & 0
              \end{pmatrix}
  \end{equation}
where $*(a) = a^*$ is the adjoint operation.
Therefore, the action of $\Theta$ on the generators of $\mathbb{T}_\theta^2$ is given by
 \begin{equation}
  \Theta \begin{pmatrix}
          u \\
          v
        \end{pmatrix} = \begin{pmatrix}
                            0 & * \\
                            - * & 0
                         \end{pmatrix} \begin{pmatrix}
                                             u \\
                                             v
                                             \end{pmatrix}  = \begin{pmatrix}
                                                                 v^* \\
                                                                 -u^*
                                                                 \end{pmatrix}
 \end{equation}
 More explicitly, the action of $\Theta$ on the generators is
 \begin{equation}
    \Theta(u) = v^*, \quad \Theta(v) = - u^*, \quad \Theta(u^*) = v, \quad \Theta(v^*) = -u
 \end{equation}

 The action of $\Theta$ respects the order of a product since we are dealing with a fermionic system, for example,
 \begin{equation}
   \Theta(uv) = \Theta(u) \Theta(v) = v^*(-u^*) = -v^*u^*
 \end{equation}
 As a consequence, this action is compatible with the relation $uv = e^{2\pi i \theta} vu$.
 
 If $x= i e^{\pi i \theta}  uv^*, y = i e^{-\pi i \theta} vu^*$, then $x, y$ are two unitary operators invariant under $\Theta$, i.e., 
   \begin{equation}
       \Theta(x) = x, \quad \Theta(y) =y
   \end{equation}

 \begin{defn}
    The fixed point algebra of the time reversal symmetry (represented by $\Theta$), denoted by $(\mathbb{T}^2_\theta)^\Theta$, is the  universal $C^*$-algebra generated by two unitary operators,
 \begin{equation}
    x= i e^{\pi i \theta}  uv^*, \quad y = i e^{-\pi i \theta} vu^*
 \end{equation}

 \end{defn} 
 By the relation $x^* = -y$, as a $C^*$-algebra, the fixed point algebra $(\mathbb{T}^2_\theta)^\Theta$ is actually generated by the unitary operator $x$. In addition, 
 $(\mathbb{T}^2_\theta)^\Theta$ is a real $C^*$-algebra with respect to the real structure $\Theta$.

 \begin{prop}
    The fixed point algebra of the time reversal symmetry 
    in $\mathbb{T}^2_\theta$ is $*$-isomorphic to the algebra of real continuous functions on the circle,
 \begin{equation}
    (\mathbb{T}^2_\theta)^\Theta   \cong C_R(\mathbb{S}^1)
 \end{equation}
 \end{prop} 

     The real K-theory of the fixed point algebra $(\mathbb{T}^2_\theta)^\Theta $  can be computed as follows,
     \begin{equation}
        KO_i[(\mathbb{T}^2_\theta)^\Theta] \cong   KO_i[C_R(\mathbb{S}^1) ]  \cong KR^{-i}(\mathbb{S}^1) = KO^{-i}(pt) \oplus KO^{-i+1}(pt)
     \end{equation}
 In particular, we have
\begin{equation}
    \widetilde{KO}_{1}[(\mathbb{T}^2_\theta)^\Theta] = \mathbb{Z}_2  % \quad  {KO}_{2}[(\mathbb{T}^2_\theta)^\Theta] = \mathbb{Z}_2 \oplus   \mathbb{Z}_2 
\end{equation}
 which is the receptacle of the topological $\mathbb{Z}_2$ invariant from the boundary K-theory if the fixed point algebra is viewed as the boundary.

Let us look at the geometry  of the fixed point algebra $(\mathbb{T}^2_\theta)^\Theta$.
From the concrete form of the generator $x = i e^{\pi i \theta}  uv^*$, it creates a virtual ``channel'' %(due to the imaginary unit ``i'')  
between $u$ and $v^*$. If one thinks of the relation $uv =  e^{2\pi i \theta} vu$ as a braiding rule, say to rotate $u$ about $v$, then one gets a phase $ e^{2\pi i \theta}$ after a twist. 
Along the same logic, $ e^{\pi i \theta}$ in $x$ is viewed as a half-twist. The same analysis applies to the other generator $y =i e^{-\pi i \theta} vu^*$,
which creates another channel from $v$ to $u^*$ in the opposite direction. Putting $x$ and $y$ (or the overlapping channels) together, they creates a virtual double cone with a crossing as in a Majorana zero mode (or a Dirac cone).
However, one channel is enough to characterize
the geometry since the opposite channel is easily recovered by $x^* = -y$. 

If we define a matrix by putting $x$ and $y$ off-diagonally, i.e.,
$ \begin{pmatrix}
  0 & y \\
  x & 0
\end{pmatrix} = \begin{pmatrix} 
                 0 & -x^* \\
                 x & 0 
                 \end{pmatrix}$,
then it is analogous to the skew-adjoint operator $\tilde{D} = \begin{pmatrix}
                                                                 0 & -D \\
                                                                 D & 0
                                                               \end{pmatrix}$. 
Hence the reduction from the pair $(x, y)$ to $x$ reminds us of the reduction from $\tilde{D}$ to $D$.
If the skew-adjoint operator  $\tilde{D}$ is indexed by the fixed point $X^\tau$, 
then the  index of  $\tilde{D}$ lives in $KO^{-2}(X^\tau)$.  Corresponding to $D$, $x$ induces the generator of $  \widetilde{KO}_{1}[(\mathbb{T}^2_\theta)^\Theta]$, 
where $(\mathbb{T}^2_\theta)^\Theta$ is the noncommutative generalization of $X^\tau$.

 \section{Noncommutative Kane--Mele invariant}
 
 The Kane--Mele invariant can be understood geometrically as the comparison between orientations of the
 Pfaffian and determinant line bundles over the fixed points. If the determinant line bundle is normalized to be a trivial bundle, then
 the Kane--Mele invariant carries information about the orientation of the Pfaffian line bundle.
 We will generalize the Kane--Mele invariant  onto the noncommutative  2-torus.

  Quillen's construction  of  the determinant line bundle on Riemann surfaces  can be generalized to the noncommutative 2-torus \cite{FGK14}.
  Let $\mathscr{F}(\mathcal{H})$ be the space of Fredholm operators over a separable infinite dimensional complex Hilbert space $\mathcal{H}$, 
  the determinant line bundle $\pi: Det \rightarrow \mathscr{F}(\mathcal{H}) $ can be defined so that for any Fredholm operator $T \in \mathscr{F}(\mathcal{H})$,
     \begin{equation}
            Det_T  = \wedge^{max}(ker\,T)^* \otimes \wedge^{max}(ker\,T)
     \end{equation}

   Assume that the single particle Hamiltonian $H$ is parametrized by the noncommutative 2-torus, that is, $H(a)$ is a family of self-adjoint Fredholm operators  
   for $a \in \mathbb{T}^2_\theta$.
    Then the effective Hamiltonian
   \begin{equation}
       \tilde{H}(a) = \begin{pmatrix}
                                                                                0 & \Theta H(a) \Theta^* \\
                                                                                H(a) & 0
                                                                               \end{pmatrix}
   \end{equation}
   is a family of Fredholm operators depending on $a  \in \mathbb{T}^2_\theta $. 
   
   \begin{examp}
    A concrete example of $H$ in our mind is $H(a) = \delta_1 + \delta_2 + a$, where $a \in (\mathbb{T}^2_{\theta})_{sa}^\times$ is a self-adjoint invertible element, 
    and $\delta_i$ $(i = 1, 2)$ are the canonical deviations on $\mathbb{T}^2_\theta$ such that
    \begin{equation}
         \delta_1(u) = u, \quad \delta_1(v) = 0, \quad  \delta_2(u) = 0, \quad \delta_2(v) = v
    \end{equation}

   \end{examp}
  
   So the determinant line bundle of $\tilde{H}$ over the noncommutative 2-torus can be  defined as the pullback line bundle
   $\mathcal{L}( \tilde{H}) := \tilde{H}^*Det$ such that
   \begin{equation}
       \mathcal{L}_{a}( \tilde{H}) = \wedge^{max}(ker H(a))^* \otimes \wedge^{max}(ker \Theta H(a)\Theta^*)
   \end{equation}
   where $(ker H(a))^* $ is the dual vector space of $ker H(a)$. $\mathcal{L}( \tilde{H})$ is the pullback bundle in  the commutative diagram,
   \begin{equation}
       \xymatrix{
       \mathcal{L}( \tilde{H}) \ar[d]^\pi \ar[r] & Det \ar[d]^\pi\\
       \mathbb{T}^2_\theta \ar[r]^{\tilde{H}} &\mathscr{F}(\mathcal{H})}
   \end{equation}
   When restricted to the fixed point algebra, by assumption, the self-adjoint Fredholm operator $H(a)$ is approximated by a Dirac operator $D(b)$ for $b \in (\mathbb{T}^2_\theta)^\Theta$,
   and the effective Hamiltonian $\tilde{H}$ becomes the skew-adjoint operator 
   \begin{equation}
       \tilde{D}(b) = \begin{pmatrix}
                    0 & -D(b) \\
                    D(b) & 0
                  \end{pmatrix}, \quad b \in (\mathbb{T}^2_\theta)^\Theta
   \end{equation} 
   Since $ker D(b) = ker (- D(b)) $ for each $b \in (\mathbb{T}^2_\theta)^\Theta$, the restricted bundle  $  \mathcal{L}( \tilde{D}) = \mathcal{L}( \tilde{H})|_{(\mathbb{T}^2_\theta)^\Theta}  $ 
   is  a trivial complex line bundle, i.e., 
   \begin{equation}
       \mathcal{L}( \tilde{D})  \cong {(\mathbb{T}^2_\theta)^\Theta} \times \mathbb{C}
   \end{equation}

   \begin{examp}
   If $H(a) = \delta_1 + \delta_2 + a$, $a \in (\mathbb{T}^2_{\theta})_{sa}^\times$,  then the Dirac operator $D(b) = \delta_1 + \delta_2 + b$,  $b \in (\mathbb{T}^2_\theta)^\Theta$. 
   $H(a)$ can be approximated by $D(b)$ if the norm of the difference between $a$ and $b$ is infinitesimal,
   i.e., $|| a - b || < \varepsilon$.     
   \end{examp}
       
 Consider Fredholm operators acting on a real Hilbert space $\mathscr{F}_\mathbb{R} := \mathscr{F}(\mathcal{H}_\mathbb{R})$,
 and let $\mathscr{F}^{sa}_\mathbb{R} \subset \mathscr{F}_\mathbb{R}$ be the subspace of self-adjoint Fredholm operators,
    which is a classifying space of $KO_{1}$, i.e., $KO_{1}(A) \cong [A, \mathscr{F}^{sa}_\mathbb{R}]$ for a real $C^*$-algebra $A$.
    Similar to the determinant line bundle, one defines the Pfaffian line bundle as $Pf \rightarrow \mathscr{F}^{sa}_\mathbb{R}$ such that
   for any $T \in \mathscr{F}^{sa}_\mathbb{R}$,      
   \begin{equation}
     Pf_T  = \wedge^{max}ker\,T
   \end{equation}
  If one complexifies the underlying real Hilbert space, then one has the relation between the Pfaffian and determinant line bundles,
     \begin{equation}
      Pf \otimes Pf \cong Det
     \end{equation}
  Or equivalently, for a real self-adjoint Fredholm operator  $T \in \mathscr{F}^{sa}_\mathbb{R}$, the fibers at $T$ can be identified as a vector space,
    \begin{equation}
     (Pf_T)^* \otimes Pf_T \cong Det_T 
    \end{equation}

  For any real element $b \in (\mathbb{T}^2_\theta)^\Theta $, 
  the Dirac operator $D(b)$ is a real self-adjoint Fredholm operator.  
  The Pfaffian line bundle over the fixed point algebra is defined similarly as the pullback line bundle  
   $\mathcal{L}^{1/2}(D) := D^*Pf$, 
   \begin{equation}
     \xymatrix{
     \mathcal{L}^{1/2}(D) \ar[d]^\pi \ar[r] & Pf \ar[d]^\pi\\
     (\mathbb{T}^2_\theta)^\Theta \ar[r]^{D} &\mathscr{F}^{sa}_\mathbb{R}}
   \end{equation}
    with a typical fiber 
   \begin{equation}
       \mathcal{L}^{1/2}_{b}(D) = \wedge^{max}ker D(b)
   \end{equation}
 As a consequence, when the determinant line bundle $\mathcal{L}(\tilde{H})$ is restricted to the fixed point algebra $(\mathbb{T}^2_\theta)^\Theta$, i.e., $ \mathcal{L}(\tilde{D})$,
 one obtains the relation 
 \begin{equation} \label{relPfDet}
     \mathcal{L}^{1/2}(D) \otimes \mathcal{L}^{1/2}(D) \cong  \mathcal{L}(\tilde{D}) = \mathcal{L}(\tilde{H})|_{(\mathbb{T}^2_\theta)^\Theta} 
 \end{equation}

    The  determinant line bundle restricted to the fixed point algebra $\mathcal{L}(\tilde{D})$ is a  trivial line bundle, 
    which has a canonical trivialization (or global section) $s$ such that $s = 1$ on the open subset of invertible operators in $(\mathbb{T}^2_\theta)^\Theta$, 
    denoted by $(\mathbb{T}^2_\theta)^{\Theta, \times}$.
    From the relation between the determinant and Pfaffian line bundles \eqref{relPfDet},
    there exists a section $ \varrho$ of the restricted Pfaffian line bundle $ \mathcal{L}^{1/2}(D)|_{(\mathbb{T}^2_\theta)^{\Theta, \times}}$ 
    such that $\varrho^2 = s = 1 $.
    
    From the isomorphism $(\mathbb{T}^2_\theta)^\Theta \cong C_R(\mathbb{S}^1)$, the invertible open set  
    $(\mathbb{T}^2_\theta)^{\Theta, \times}$ can be identified with 
    \begin{equation}
           C_R(\mathbb{S}^1)^\times = \{ f \in C_R(\mathbb{S}^1) \,\,s.t.\,\, f(\vartheta) \neq 0 \quad \forall \,\, \vartheta \in \mathbb{S}^1 \}
    \end{equation}
    The range of  $f \in C_R(\mathbb{S}^1)^\times $ is $\mathbb{C}^\times$, after the normalization $\mathbb{C}^\times \mapsto \mathbb{S}^1$, 
    $f$ is viewed as a continuous function $f : \mathbb{S}^1 \rightarrow \mathbb{S}^1$
    such that $\overline{f(\vartheta)} = f(-\vartheta)$. For example, one has $f_n: \vartheta \mapsto e^{in\vartheta}$ (or $f_n(z) = z^n,\, z = e^{i \vartheta}$)
    as an element in $C_R(\mathbb{S}^1)^\times $.
    In fact, $\{z^n, \bar{z}^n\}$ forms  an orthonormal basis  for $C_R(\mathbb{S}^1)^\times $, by the relation $(\mathbb{T}^2_\theta)^{\Theta, \times} \cong C_R(\mathbb{S}^1)^\times$, 
    the generator $x \in (\mathbb{T}^2_\theta)^{\Theta, \times}$ (resp. $y$) is identified with $z \in C_R(\mathbb{S}^1)^\times$ (resp. $\bar{z}$). 
    On the other hand, the Gelfand transform of the fixed point algebra $(\mathbb{T}^2_\theta)^\Theta \cong C^*_\mathbb{R}(\mathbb{Z})$
    can be identified with the Fourier transform on the circle. 
    
    The fixed point algebra $(\mathbb{T}^2_\theta)^\Theta$ is generated by two unitaries $x$ and $y$ (with $x^* = -y$).
     We only care about the orientation of the real Pfaffian line bundle $\pi: \mathcal{L}^{1/2}(D) \rightarrow (\mathbb{T}^2_\theta)^\Theta \cong C_R(\mathbb{S}^1) $, 
    so its structure group can be reduced from $GL(1, \mathbb{R}) = \mathbb{R}^\times$ to $O(1) = \mathbb{Z}_2$. Indeed, the orientation of $ \mathcal{L}^{1/2}(D)$ is essentially 
    determined by the value of $\varrho$  on the generators,
    \begin{equation}
         \varrho : \{ x, y\} \rightarrow O(1)
    \end{equation}

    \begin{defn}
       The noncommutative Kane--Mele invariant is defined by
    \begin{equation}
        \nu= \nu[\mathcal{L}^{1/2}(D)] :=  \varrho(x) \varrho(y) \in \mathbb{Z}_2
    \end{equation}
    where $x$ and $y$ are the generators of the fixed point algebra $(\mathbb{T}^2_\theta)^\Theta$ and $\varrho$ is the canonical section of the Pfaffian line bundle
    $\pi: \mathcal{L}^{1/2}(D) \rightarrow (\mathbb{T}^2_\theta)^\Theta $ such that $\varrho^2 = 1$.
    \end{defn}

    \begin{prop}
       The noncommutative Kane--Mele invariant $\nu[ \mathcal{L}^{1/2}(D)]$ tells us the orientability of the Pfaffian line bundle $\pi: \mathcal{L}^{1/2}(D) \rightarrow (\mathbb{T}^2_\theta)^\Theta $, and 
       $\nu[ \mathcal{L}^{1/2}(D)]$ is identified as an element in $\widetilde{KO}_1[(\mathbb{T}^2_\theta)^\Theta] $.
    \end{prop}

\section{Noncommutative bulk-boundary correspondence} 
  
  In this final section,  we establish the equivalence between   the    topological $\mathbb{Z}_2$ index and the  Kane--Mele invariant in the noncommutative 2-torus $\mathbb{T}^2_\theta$.
The bulk-boundary correspondence is  a map from the KR-theory of $\mathbb{T}^2_\theta$ to the KO-theory of the fixed point algebra $(\mathbb{T}^2_\theta)^\Theta$.

  The topological $\mathbb{Z}_2$ index lives in the KR-theory of the bulk, which represents the momentum space in the classical case. 
  The Kane--Mele invariant is originally defined over the set of fixed points, which is identified as the effective boundary in our treatment,
  so the Kane--Mele invariant falls into the KO-theory of the boundary. 
  The bulk-boundary correspondence is the key in  understanding the relevant index theory using bivariant K-theory (i.e., KK-theory). 
  First, the  bulk-boundary correspondence identifies the topological $\mathbb{Z}_2$ index  $ind_t(w)$ with the Kane--Mele invariant $\nu$,
   \begin{equation}\label{CSequivKM}
      \nu = (-1)^{ind_t(w)}
   \end{equation}
  where $w$ is the specific gauge of the time reversal symmetry.
  Second, the bulk-boundary correspondence can be understood as a KKR-cycle  generalizing the topological index map, 
 \begin{equation}
          KR^{-2-d}(X) \times KKR^{d}(X, X^\tau) \rightarrow    KO^{-2}(X^\tau)  
\end{equation}
 In particular, when $X = \mathbb{T}^2$, the bulk-boundary correspondence is a map,
 \begin{equation}
     {KQ}(\mathbb{T}^2) = {KR}^{-4}(\mathbb{T}^2) =   KO^{-4}(pt) \oplus KO^{-2}(pt) \rightarrow KO^{-2}(pt)
 \end{equation}

 We have to point out our proposal on the bulk-boundary correspondence is different from those in the literature, for example \cite{BCR16, BKR17, K16, MT16}, because  the effective boundary is taken as 
 the set of fixed point, which is motivated by the Kane--Mele invariant. 
 Instead of a correspondence between a $d$-dimensional bulk and a $(d-1)$-dimensional boundary,  we proposed a correspondence between
 a  $d$-dimensional bulk and a $0$-dimensional effective boundary in the classical case (i.e., $X$ vs. $X^\tau$). 
  The bulk-boundary correspondence is supposed to be a KKR-cycle connecting the KR-theory of the bulk and the KO-theory of the (effective) boundary.

  In noncommutative topology, we consider the bulk as  $\mathbb{T}^2_\theta$, which is the noncommutative function algebra generalizing $C(\mathbb{T}^2)$. 
  As a generalization of the fixed points, we consider the boundary as the fixed point algebra $(\mathbb{T}^2_\theta)^\Theta$, 
  which is a real $C^*$-subalgebra of $\mathbb{T}^2_\theta$ such that $(\mathbb{T}^2_\theta)^\Theta \cong C_R(\mathbb{S}^1)$. 
  We are interested in the orientability of the Pfaffian line bundle  $\pi: \mathcal{L}^{1/2}(D) \rightarrow (\mathbb{T}^2_\theta)^\Theta$, 
  so we look at the real K-group $KO_1[(\mathbb{T}^2_\theta)^\Theta]$ and define the noncommutative Kane--Mele invariant $\nu[\mathcal{L}^{1/2}(D)] \in \widetilde{KO}_1[(\mathbb{T}^2_\theta)^\Theta]$.
  
  On the one hand, the classical Kane--Mele invariant 
  \begin{equation}
     \nu = [ind(\tilde{D})] \in  KO^{-2}(X^\tau) \rightarrow KO^{-2}(pt) 
  \end{equation}
  where $\nu$ is interpreted as the analytical index of the skew-adjoint operator $\tilde{D}$.
  On the other hand,
  \begin{equation}
       \nu[\mathcal{L}^{1/2}(D)] \in \widetilde{KO}_1[(\mathbb{T}^2_\theta)^\Theta] = \widetilde{KR}^{-1}(\mathbb{S}^1) = KO^{-1}(pt)
  \end{equation}
  The reduction in KO-theory from $KO^{-2}$ to $KO^{-1}$ agrees with the reduction from $\tilde{D}$ to a Dirac operator $D$.
  
  Let us get back to the bulk theory, in the classical case, the topological $\mathbb{Z}_2$ index lives in ${KR}^{-4}(\mathbb{T}^2) = KQ(\mathbb{T}^2)$,
  \begin{equation}
      ind(w) \in {KR}^{-4}(\mathbb{T}^2) = KO^{-2}(pt) \oplus KO^{-4}(pt)
  \end{equation}
  By the same reduction from $\tilde{D}$ to $D$, we conclude that the noncommutative topological $\mathbb{Z}_2$ index belongs to ${KR}_{2}(\mathbb{T}^2_\theta)$,
  \begin{equation}
       ind(PwP) \in {KR}_{2}(\mathbb{T}^2_\theta)
  \end{equation}

\begin{prop}
   The $KR$-theory of the noncommutative 2-torus $\mathbb{T}^2_\theta$ is computable by the Pimsner--Voiculescu method.
   In particular, we have 
   \begin{equation}
            KR_2(\mathbb{T}^2_\theta) = KO^{-1}(pt) \oplus 2 KO^{-2}(pt) %= 3\mathbb{Z}_2
   \end{equation}
    which is the receptacle of the topological $\mathbb{Z}_2$ index from the bulk K-theory.
\end{prop}

   \begin{thm}
      The noncommutative Kane--Mele invariant and the noncommutative topological $\mathbb{Z}_2$ index are equivalent.
   \end{thm}

   The above theorem tells us that the  bulk-boundary correspondence is basically an index theorem identifying the topological index from the bulk and the analytical index from the boundary. 
   On the level of K-theory, the bulk-boundary correspondence involves a KKR-cycle  connecting the K-theories of the bulk and boundary,
   \begin{equation}
      KKR( (\mathbb{T}^2_\theta)^\Theta,  \mathbb{T}^2_\theta ) \times KR_2(\mathbb{T}^2_\theta) \rightarrow \widetilde{KO}_1 [ (\mathbb{T}^2_\theta)^\Theta]
   \end{equation}
   that identifies the $KO^{-1}(pt)$ component from both sides. 

\nocite{*}
\bibliographystyle{plain}
\bibliography{chpt}

\begin{thebibliography}{10}

\bibitem{A66}
M.~Atiyah.
\newblock K-theory and reality.
\newblock {\em Quart. J. Math.}, 17(2):367--386, 1966.

\bibitem{AS69}
M.~Atiyah and I.M. Singer.
\newblock Index theory for skew-adjoint {F}redholm operators.
\newblock {\em Inst. Hautes Etudes Sci. Publ. Math.}, 37:5--26, 1969.

\bibitem{AS71}
M.~Atiyah and I.M. Singer.
\newblock The index of elliptic operators: V.
\newblock {\em Ann. of Math.}, 93:139--149, 1971.

\bibitem{ASV13}
J.C. Avila, H.~Schulz-Baldes, and C.~Villegas-Blas.
\newblock Topological invariants of edge states for periodic two-dimensional
  models.
\newblock {\em Math. Phys., Anal. Geom.}, 16:136--170, 2013.

\bibitem{ASS94}
J.~Avron, R.~Seiler, and B.~Simon.
\newblock The charge deficiency, charge transport and comparison of dimensions.
\newblock {\em Commun. Math. Phys.}, 159:399--422, 1994.

\bibitem{BES94}
J.~Bellissard, A.~van Elst, and H.~Schulz-Baldes.
\newblock The noncommutative geometry of the quantum {H}all effect.
\newblock {\em J. Math. Phys.}, 35:5373--5451, 1994.

\bibitem{BCR16}
C.~Bourne, A.L. Carey, and A.~Rennie.
\newblock A noncommutative framework for topological insulators.
\newblock {\em Rev. Math. Phys.}, 28:1650004, 2016.

\bibitem{BKR17}
C.~Bourne, J.~Kellendonk, and A.~Rennie.
\newblock The {K}-theoretic bulk-edge correspondence for topological
  insulators.
\newblock {\em Ann. Henri Poincar{\'e}}, 18:1833, 2017.

\bibitem{CGLW13}
X.~Chen, Z.C. Gu, Z.X. Liu, and X.G. Wen.
\newblock Symmetry protected topological orders and the group cohomology of
  their symmetry group.
\newblock {\em Phys. Rev. B}, 87:155114, 2013.

\bibitem{C94}
A.~Connes.
\newblock {\em Noncommutative Geometry}.
\newblock Academic Press, 1994.

\bibitem{C95}
A.~Connes.
\newblock Noncommutative geometry and reality.
\newblock {\em J. Math. Phys.}, 36:11, 1995.

\bibitem{DFN15}
S.~{Das Sarma}, M.~Freedman, and C.~Nayak.
\newblock Majorana zero modes and topological quantum computation.
\newblock {\em Quantum Information}, 1:15001, 2015.

\bibitem{DG15}
G.~{De Nittis} and K.~Gomi.
\newblock Classification of ``{Q}uaternionic'' {B}loch-bundles: Topological
  insulators of type {AII}.
\newblock {\em Commun. Math. Phys.}, 339:1--55, 2015.

\bibitem{DS16}
G.~{De Nittis} and H.~Schulz-Baldes.
\newblock Spectral flows associated to flux tubes.
\newblock {\em Annales Henri Poincare}, 17:1--35, 2016.

\bibitem{FGK14}
A.~Fathi, A.~Ghorbanpour, and M.~Khalkhali.
\newblock The curvature of the determinant line bundle on the noncommutative
  two torus.
\newblock {\em Math. Phys. Anal. Geom.}, 20:4, 2014.

\bibitem{F87}
D.~Freed.
\newblock On determinant line bundles.
\newblock In S.~T. Yau, editor, {\em Math. aspects of string theory}. World
  Sci. Publ., 1987.

\bibitem{FM13}
D.~Freed and G.M. Moore.
\newblock Twisted equivariant matter.
\newblock {\em Annales Henri Poincar{\'e}}, 14(8):1927--2023, 2013.

\bibitem{FK06}
L.~Fu and C.~Kane.
\newblock Time reversal polarization and a $\mathbb{Z}_2$ adiabatic spin pump.
\newblock {\em Phys. Rev. B}, 74:195312, 2006.

\bibitem{FKM07}
L.~Fu, C.~Kane, and E.~Mele.
\newblock Topological insulators in three dimensions.
\newblock {\em Phys. Rev. Lett.}, 98:106803, 2007.

\bibitem{G17}
K.~Gawedzki.
\newblock 2d {F}u--{K}ane--{M}ele invariant as {W}ess--{Z}umino action of the
  sewing matrix.
\newblock {\em Lett. Math. Phys.}, 107(4):733--755, 2017.

\bibitem{G93}
E.~Getzler.
\newblock The odd {C}hern character in cyclic homology and spectral flow.
\newblock {\em Topology}, 32(3):489--507, 1993.

\bibitem{GP13}
G.M. Graf and M.~Porta.
\newblock Bulk-edge correspondence for two-dimensional topological insulators.
\newblock {\em Commun. Math. Phys.}, 324(3):851--895, 2013.

\bibitem{GS15}
J.~Grossmann and H.~Schulz-Baldes.
\newblock Index pairings in presence of symmetries with applications to
  topological insulators.
\newblock {\em Commum. Math. Phys.}, 343(2):477--513, 2015.

\bibitem{HK10}
M.Z. Hasan and C.L. Kane.
\newblock Colloquium: Topological insulators.
\newblock {\em Reviews of Modern Physics}, 82(4):3045, 2010.

\bibitem{HR00}
N.~Higson and J.~Roe.
\newblock {\em Analytic {K}-homology}.
\newblock Oxford Science Publications, 2000.

\bibitem{KM05}
C.~Kane and E.~Mele.
\newblock $\mathbb{Z}_2$ topological order and the quantum spin {H}all effect.
\newblock {\em Phys. Rev. Lett.}, 95:146802, 2005.

\bibitem{KM0501}
C.~Kane and E.~Mele.
\newblock Quantum spin {H}all effect in graphene.
\newblock {\em Phys. Rev. Lett.}, 95:226801, 2005.

\bibitem{Kar78}
Max Karoubi.
\newblock {\em {$K$}-theory}.
\newblock Classics in Mathematics. Springer-Verlag, Berlin, 1978.

\bibitem{KK16}
H.~Katsura and T.~Koma.
\newblock The $\mathbb{Z}_2$ index of disordered topological insulators with
  time reversal symmetry.
\newblock {\em J. Math. Phys.}, 57:021903, 2016.

\bibitem{KLW1501}
R.~Kaufmann, D.~Li, and B.~Wehefritz-Kaufmann.
\newblock Topological insulators and {K}-theory.
\newblock 2015.
\newblock arXiv: 1510.08001.

\bibitem{KLW1601}
R.~Kaufmann, D.~Li, and B.~Wehefritz-Kaufmann.
\newblock Noncommutative topological $\mathbb{Z}_2$ invariant.
\newblock 2016.
\newblock arXiv: 1605.09470.

\bibitem{KLW15}
R.~Kaufmann, D.~Li, and B.~Wehefritz-Kaufmann.
\newblock Notes on topological insulators.
\newblock {\em Rev. Math. Phys.}, 28(10):1630003, 2016.

\bibitem{KLW16}
R.~Kaufmann, D.~Li, and B.~Wehefritz-Kaufmann.
\newblock The {S}tiefel--{W}hitney theory of topological insulators.
\newblock 2016.
\newblock arXiv: 1604.02792.

\bibitem{Kell16}
J.~Kellendonk.
\newblock Cyclic cohomology for graded ${C}^{\ast,r}$-algebras and its pairings
  with van daele {K}-theory.
\newblock 2016.
\newblock arXiv: 1607.08465.

\bibitem{K15}
J.~Kellendonk.
\newblock On the ${C}^*$-algebraic approach to topological phases for
  insulators.
\newblock {\em Annales Henri Poincar{\'e}}, 18:2251, 2017.

\bibitem{K09}
A.~Kitaev.
\newblock Periodic table for topological insulators and superconductors.
\newblock {\em AIP Conf. Proc.}, 1134:22--30, 2009.

\bibitem{K16}
Y.~Kubota.
\newblock Controlled topological phases and bulk-edge correspondence.
\newblock {\em Commun. Math. Phys.}, pages 1--33, 2016.

\bibitem{LM90}
H.~B. Lawson and M.~Michelsohn.
\newblock {\em Spin geometry}.
\newblock Princeton University Press, 1990.

\bibitem{L13}
D.~Li.
\newblock Noncommutative chern-simons theory on the quantum 3-sphere.
\newblock 2013.
\newblock arXiv: 1310.7233.

\bibitem{L17}
D.~Li.
\newblock Bott--kitaev periodic table and index theory.
\newblock 2017.
\newblock arXiv: 1710.01426.

\bibitem{L88}
J.~Lott.
\newblock Real anomalies.
\newblock {\em J. Math. Phys.}, 29:1455--1464, 1988.

\bibitem{MT16}
V.~Mathai and G.C. Thiang.
\newblock T-duality simplifies bulk-boundary correspondence.
\newblock {\em Commun. Math. Phys.}, 345(2):675--701, 2016.

\bibitem{MB07}
J.~Moore and L.~Balents.
\newblock Topological invariants of time-reversal-invariant band structures.
\newblock {\em Phys. Rev. B.}, 75:121306, 2007.

\bibitem{N03}
M.~Nakahara.
\newblock {\em Geometry, Topology and Physics}.
\newblock CRC Press, 2003.

\bibitem{P96}
J.~Phillips.
\newblock Self-adjoint {F}redholm operators and spectral flow.
\newblock {\em Canad. Math. Bull.}, 39:460--467, 1996.

\bibitem{PV80}
M.~Pimsner and D.~Voiculescu.
\newblock Exact sequences for {K}-groups and {E}xt-groups of certain
  cross-product ${C}^*$-algebras.
\newblock {\em J. Operator Theory}, 4(1):93--118, 1980.

\bibitem{P11}
E.~Prodan.
\newblock Disordered topological insulators: a non-commutative geometry
  perspective.
\newblock {\em J. Phys. A: Math. Theor.}, 44(11):113001, 2011.

\bibitem{PS16}
E.~Prodan and H.~Schulz-Baldes.
\newblock {\em Bulk and boundary invariants for complex topological insulators:
  From {K}-theory to physics}.
\newblock Springer, Berlin, 2016.

\bibitem{QHZ08}
X.L. Qi, T.~Hughes, and S.C. Zhang.
\newblock Topological field theory of time-reversal invariant insulators.
\newblock {\em Phys. Rev. B}, 78:195424, 2008.

\bibitem{QZ11}
X.L. Qi and S.C. Zhang.
\newblock Topological insulators and superconductors.
\newblock {\em Rev. Mod. Phys.}, 83:1057--1111, 2011.

\bibitem{R15}
J.~Rosenberg.
\newblock Structure and applications of real ${C}^*$-algebras.
\newblock In R.~Doran and E.~Park, editors, {\em Operator Algebras and their
  Applications}, volume 671 of {\em Contemporary Math.}, pages 235--258. Amer.
  Math. Soc., Providence, RI, 2016.

\bibitem{SRFL09}
A.~Schnyder, S.~Ryu, A.~Furusaki, and A.~Ludwig.
\newblock Classification of topological insulators and superconductors.
\newblock {\em AIP Conf. Proc.}, 1134:10--21, 2009.

\bibitem{S16}
H.~Schulz-Baldes.
\newblock Topological insulators from the perspective of non-commutative
  geometry and index theory.
\newblock {\em Jahresber Dtsch Math-Ver}, 118:247--273, 2016.

\bibitem{S13}
S.Q. Shen.
\newblock {\em Topological Insulators: {D}irac Equation in Condensed Matters}.
\newblock Springer, 2013.

\bibitem{T16}
G.C. Thiang.
\newblock On the {K}-theoretic classification of topological phases of matter.
\newblock {\em Annales Henri Poincare}, 17(4):757--794, 2016.

\bibitem{VFG01}
J.~C. Varilly, H.~Figueroa, and J.~M. Gracia-Bondia.
\newblock {\em Elements of Noncommutative Geometry}.
\newblock Birkh{\"a}user, 2001.

\bibitem{WQZ10}
Z.~Wang, X.L. Qi, and S.C. Zhang.
\newblock Equivalent topological invariants of topological insulators.
\newblock {\em New J. Phys.}, 12:065007, 2010.

\bibitem{W52}
H.~Weyl.
\newblock {\em Symmetry}.
\newblock Princeton University Press, 1952.

\bibitem{W82}
E.~Witten.
\newblock An {SU}(2) anomaly.
\newblock {\em Phys. Lett. B}, 117(5):324--328, 1982.

\bibitem{W16}
E.~Witten.
\newblock Fermion path integrals and topological phases.
\newblock {\em Rev. Mod. Phys.}, 88:035001, 2016.

\end{thebibliography}

 \end{document}